    \documentclass[sigconf, authorversion, nonacm]{acmart}

\usepackage{balance}
\usepackage{stfloats}

\AtBeginDocument{%
  \providecommand\BibTeX{{%
    \normalfont B\kern-0.5em{\scshape i\kern-0.25em b}\kern-0.8em\TeX}}}

\copyrightyear{2026}
\acmYear{2026}
\setcopyright{rightsretained}\acmConference[]{Preprint}{2026}{ }
\acmBooktitle{ }
\acmDOI{ }
\acmISBN{}

\begin{document}

\title{Protosampling: Enabling Free-Form Convergence of Sampling and Prototyping through Canvas-Driven Visual AI Generation}


\author{Alicia Guo}
\affiliation{
    \institution{Autodesk Research}
    \city{Toronto}
    \state{ON}
    \country{Canada}}
\email{alicia.guo@autodesk.com}

\author{David Ledo}
\affiliation{
    \institution{Autodesk Research}
    \city{Toronto}
    \state{ON}
    \country{Canada}}
\email{david.ledo@autodesk.com}

\author{George Fitzmaurice}
\affiliation{
    \institution{Autodesk Research}
    \city{Toronto}
    \state{ON}
    \country{Canada}}
\email{george.fitzmaurice@autodesk.com}

\author{Fraser Anderson}
\affiliation{
    \institution{Autodesk Research}
    \city{Toronto}
    \state{ON}
    \country{Canada}}
\email{fraser.anderson@autodesk.com}


\newcommand{\alicia}[1]{{\color{orange}\bf{Alicia: #1}\normalfont}}
\newcommand{\david}[1]{{\color{purple}\bf{David: #1}\normalfont}}
\newcommand{\fraser}[1]{{\color{cyan}\bf{Fraser: #1}\normalfont}}
\newcommand{\revision}[1]{\textcolor{black}{#1}}

\definecolor{model}{HTML}{EF4340}
\definecolor{clip}{HTML}{F1C12B}
\definecolor{ref}{HTML}{26B0BF}

\newcommand{\detailtexcount}[1]{%
  \immediate\write18{texcount -merge -sum -q #1.tex output.bbl > #1.wcdetail }%
  \verbatiminput{#1.wcdetail}%
}

\def\myurl{\hfil\penalty 100 \hfilneg \hbox}

\def \revised #1{\textcolor{black}{#1}}

\begin{abstract} 
    As an emergent process, creativity relies on explorations via sampling and prototyping for problem construction. These activities compile knowledge, provide a context enveloping the solution, and answer questions. With Generative AI, practitioners can go beyond sampling existing media towards instantly generating and remixing new ones. We refer to this convergence as 'protosampling'. Using existing literature we ground a definition for protosampling and operationalize it through Atelier, a canvas-like system that leverages a variety of generative image and video models for visual creation. Atelier: (1) blends the spaces for thinking and creation, where both references and generated assets co-exist in one space, (2) provides various encapsulated technical workflows that focus on the activity at hand, and (3) enables navigating emergence through interactive visualizations, smart search, and collections. Protosampling as a lens reframes creative work to emphasize the process itself and how seemingly disjointed thoughts can tightly interweave into a final solution.
\end{abstract}

\ccsdesc[500]{Human-centered computing~Human computer interaction (HCI)}

\keywords{Creativity Support Tools, Generative AI and Creativity, AI images and Videos, Sampling, Prototyping, Canvas-Based Interactions}


 \begin{teaserfigure}
   \begin{center}
          \vspace{-5mm}
       \includegraphics[width=1\textwidth]{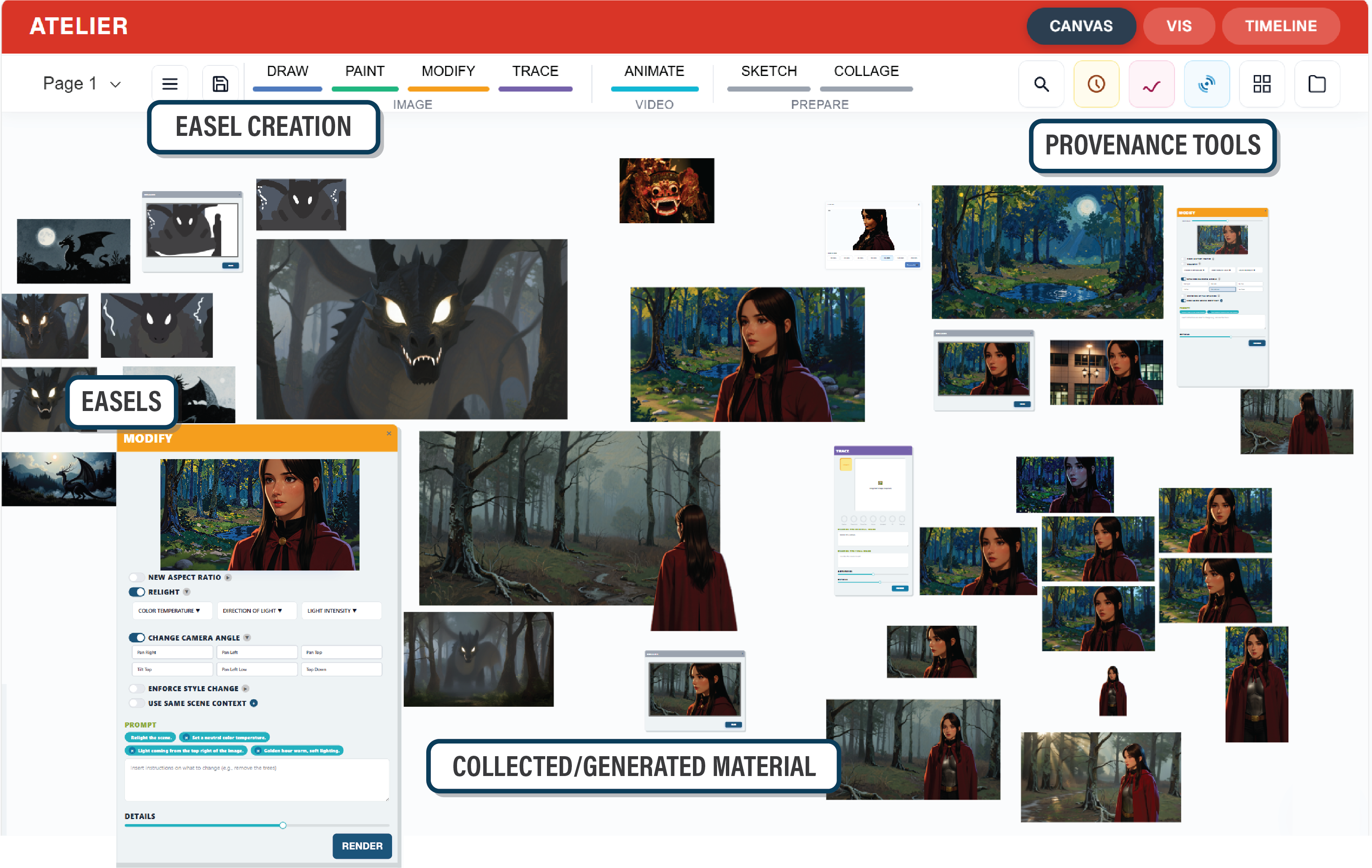}
       \vspace{-5mm}
       \caption{Atelier is a system that operationalizes protosampling - the practice of tightening sampling and prototyping for creating AI visuals. It enables a co-existence between collected and generated assets, provides a set of 'easels' that encapsulate complex workflows, and fosters collection and reflection through provenance.}
       \Description{
       Atelier is a system that operationalizes protosampling - the practice of tightening sampling and prototyping for creating AI visuals. It enables a co-existence between collected and generated assets, provides a set of 'easels' that encapsulate complex workflows, and fosters collection and reflection through provenance.
       }
       \label{fig:teaser}
    \end{center}
 \end{teaserfigure}


\maketitle

\section{Introduction}
The accelerated growth of Generative AI has enabled new capabilities for generating rich media (e.g., images, videos, 3D models, audio) using natural language as its primary means. Over time, additional mechanisms have enabled more expressiveness and control by (1) providing multi-modal inputs, such as using images to support the structure of a new generated image \cite{zhang2023controlnet}, and (2) exposing relevant parameters that can affect the resulting generation. Generative AI interfaces exist on a broad spectrum catering to different audiences. On one end, simplified chat-like systems leverage conversational language to provide quick solutions. Next, there are tools that expose some degree of inputs and controls. For instance, with MidJourney \cite{midjourney} one can type a prompt, select an aspect ratio, provide a style reference, etc. Lastly, at the very end of the spectrum, there are programmatic tools that offer the full gamut of models, inputs, and parameter settings all while requiring specialized expertise. One popular system is ComfyUI \cite{ComfyUI}, which offers a node-based programming approach to combine and control multiple models at a given time. These systems encapsulate key atomic operations into nodes that can interconnect multiple models, algorithms, and even online services. While rich and expressive, this paradigm focuses primarily on procedural thinking to solve a problem, and requires one to source a myriad of models to select the most adequate ones for the task at hand. 

This spectrum leaves out a considerable gap for creative practitioners who could benefit from a balance of control while still working through a medium that enables them to best express their intent. This means bringing their media and their process center stage, and making tools accessible when and where they are needed. By focusing on the creative process, it becomes possible to harness a latent quality of generative tools -- the ability to combine a wide variety of materials and thinking, use them to generate new materials, and continue engaging in a vast emergent recombination \cite{cross1997descriptive}, a true conversation with the creative problem at hand \cite{schon2017reflective}. This posits an interweaving of two key creative activities: sampling and prototyping. Looking at these two activities in tandem enables rethinking the creative medium as one where practitioners use the creative problem at hand as a lens to look at the world and curate information and actively transform it into partial solutions that over time inform the final result.

To explore how to support the interplay between sampling and prototyping, we designed Atelier (Figure \ref{fig:teaser}, a canvas-like interface of content, both source material, and generated, and a set of activity-centric widgets offering Generative AI operations with a set of carefully selected inputs and parameters. Atelier provides mechanisms to engage with the content via collections, search, history and interactive visualizations that enable reflection and reinterpretation of the body of work towards what might be the emergent traits within the process. \revision{An extended 4-hour first-use study with 5 creative professionals further demonstrates the expressiveness and how it might be used in real-world scenarios}. Specifically, this paper proposes the following contributions:

\begin{enumerate}
    \item A characterization of 'Protosampling' -- a lens that treats sampling and prototyping as a joint activity rather than separate stages of the creative process.
    \item Atelier, a canvas-like system that operationalizes Protosampling to support the generation of media, with a particular focus on visuals (images and video).
    \item A set of abstractions and novel workflows for generating AI images and videos while focusing on creative controllability, demonstrated via a set of usage scenarios.
\end{enumerate}
\begin{figure*}
    \centering
    \includegraphics[width = 1.0\textwidth]{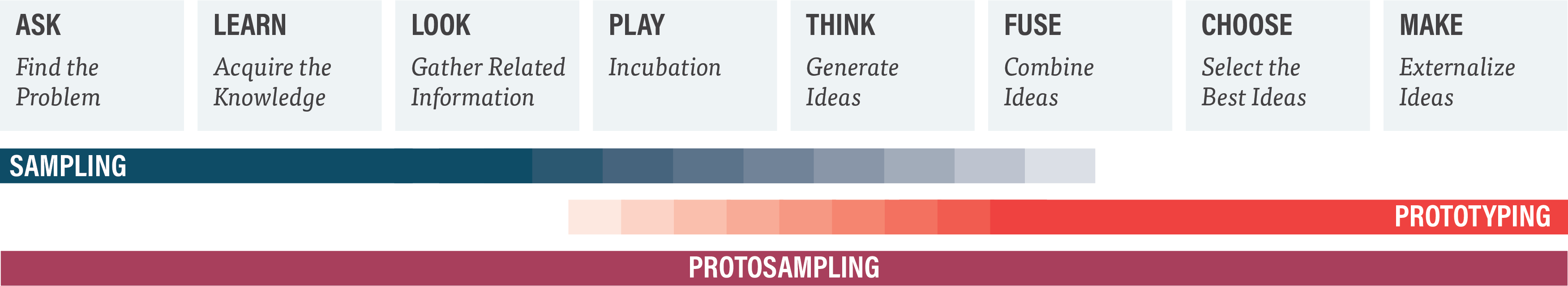}
     \caption{Overview of the creative process according to Sawyer \cite{sawyer2024explaining}, situating sampling, prototyping and proposing Protosampling as the bridge between these two activities.}
    \label{fig:ProtosamplingInProcess}
    \Description{}
\end{figure*}

\section{Protosampling}
The creative process is often described as a discrete set of stages practitioners undergo when solving a creative problem \cite{sawyer2024explaining, kaufman2016creativity}. \revision{There are myriads of representations of creativity and design processes \cite{Dubberly, kochanowska2021double, cross2011understanding, sawyer2024explaining, kaufman2016creativity} all breaking the process into discrete steps or stages from problem to solution. However, Sawyer's model \cite{sawyer2024explaining} is one of the few acknowledging that creativity does not necessarily happen in a given order, rather practitioners carry out different `disciplines'. Here, the stages are a simplification that emphasize the} most common primary activity within the process (Figure \ref{fig:ProtosamplingInProcess}). The activities are: ask (find the problem), learn (acquire the knowledge), look (gather relevant information), play (incubation), think (generate ideas), fuse (combine ideas), choose (select ideas) and make (externalize ideas) \cite{sawyer2024explaining}. Sawyer's model highlights that creative work is highly malleable and dependent on the activity and the person carrying it out. Many activities conducted in the creative process cover multiple of these stages or disciplines. Two common activities are \textit{sampling} and \textit{prototyping}. 

\textbf{\textit{Sampling}} refers to \textit{the act of collecting, organizing and transforming materials} \cite{stemasov2023immersivesampling} beyond just gathering materials. Creative practitioners engage in sampling the real world \cite{greenbergSketchingUserExperiences2011, luceroFramingAligningParadoxing2012}, continuously collecting materials and information, ranging from inspiration to relevant bits to the problem at hand. Sampling is a form of \textit{`opportunistic assimilation'}: as people go about their day, they encounter objects or situations relevant to an unsolved problem \cite{lockhart1988conceptual}. Under this premise, the creative problem becomes a lens from which to look at the world, creators look at information in their environment to link new information and integrate it into existing problems and tasks. \revision{For example, a musician might capture a soundscape from an environment, trim the audio, remove the noise, and use it as a beat in a piece, or perhaps find a chord progression from a song they like which inspires a composition. A graphic designer might keep a postcard because they like the colours. Ultimately, materials are collected, curated, in some cases formalized (e.g., mood boards) and then used.} As part of creative problem-solving, Stemasov et al. \cite{stemasov2023immersivesampling} characterize sampling from past literature as an act that is open-ended, deliberate, and helps structure thinking. Sampling provides context around creative work, synthesizes existing collections into new designs, offers potential raw materials, collects design decisions, and provides triggers to reinterpret the problem.

\textbf{\textit{Prototyping}}, on the other hand, focuses on actual creation to better understand a problem. Prototypes are seen as \textit{purposeful manifestations of design ideas that traverse a design space leading to meaningful knowledge about some aspects of a final design} \cite{lim2008anatomyofprototypes}. Prototyping can serve different roles, \revision{such as exploration, experimentation, or evolution \cite{floyd1984systematic}}. Of particular interest is the ability to explore design spaces \revision{-- a "fuzzy" boundary between exploration and experimentation \cite{floyd1984systematic}}. While Buxton distinguishes \textit{sketches} from \textit{prototypes} \cite{buxton2010sketching}, much of the design literature \cite{logan2013creativity, lim2008anatomyofprototypes, cross2011understanding, goel1992structure} advocates for prototyping the same way Buxton frames sketching: prototypes enable exploring design spaces, test out alternatives, and inform rationales \cite{lim2008anatomyofprototypes}. In fact, Stolterman argues that sketching is a rigorous approach to explore and iterate ideas. Prototypes are strong \textit{because} they are incomplete solutions, acting as \textit{`filters'} that examine the quality of an idea and answer a specific question without having to create a the final design \cite{lim2008anatomyofprototypes}. Existing work often frames prototyping in terms of exploring specific areas of a design such as function, structure, behaviour, look and feel, role, implementation, etc. \cite{cross1997descriptive, lim2008anatomyofprototypes, gero1994computational}. Lim et al. articulate how prototypes examine one or more properties of the design space \cite{lim2008anatomyofprototypes}, for example exploring appearance by tackling properties such as colour or shape. While the term `prototyping' is often used in design contexts, it applies to creative practice more broadly. Prototyping in animation might resemble different activities serving various purposes: \textit{storyboards} examine camera angles and how the story translates to visuals, and \textit{animatics} explore the timing. To create the storyboard, an artist might make small \textit{thumbnails} to see the composition. This is why Logan and Smithers \cite{logan2013creativity} argue that prototypes are all interlinked with each other.

In a broad sense, both sampling and prototyping are exploratory activities enable practitioners to engage in \textbf{\textit{"problem construction"}} \cite{kaufman2016creativity}, a process in which open-ended elements of the problem at hand are tightened and reframed to drive the solution. Cross \cite{cross2011understanding} describes problem construction as a process that begins with ambitious high level goals, followed by periods of intense activity that are followed by reflective contemplation until a sudden insight compounds the new understanding to redefine the problem. The problem and solution co-evolve.

\paragraph{\textbf{The Impact of Generative AI}} While there is still limited understanding on the implications of Generative AI to the creative process, the reality is that it has fundamentally changed it. Media such as text, images, videos, 3D models, music, etc. can all be instantiated rapidly. Materials can be remixed and reused in new generations. Furthermore, this process means many more partial solutions get created in shorter bursts of time. For example, Ledo's work \cite{ledo2025generative} shows how generating an seconds-long animation required over 600 generations. This type of AI-driven creation is bound to encompass snapshots of the process, with each generation embodying parts of the understanding of the problem at that given point. \textit{Sampling has become more like prototyping}, as assets are not only collected but also created, \textit{and prototyping is becoming more like sampling}, as both collected and generated materials can be combined and reused. Seeing as this loop might be becoming tighter, we see value in bridging these two concepts together: \textit{Protosampling}.

\begin{figure*}
    \centering
    \includegraphics[width = 1.0\textwidth]{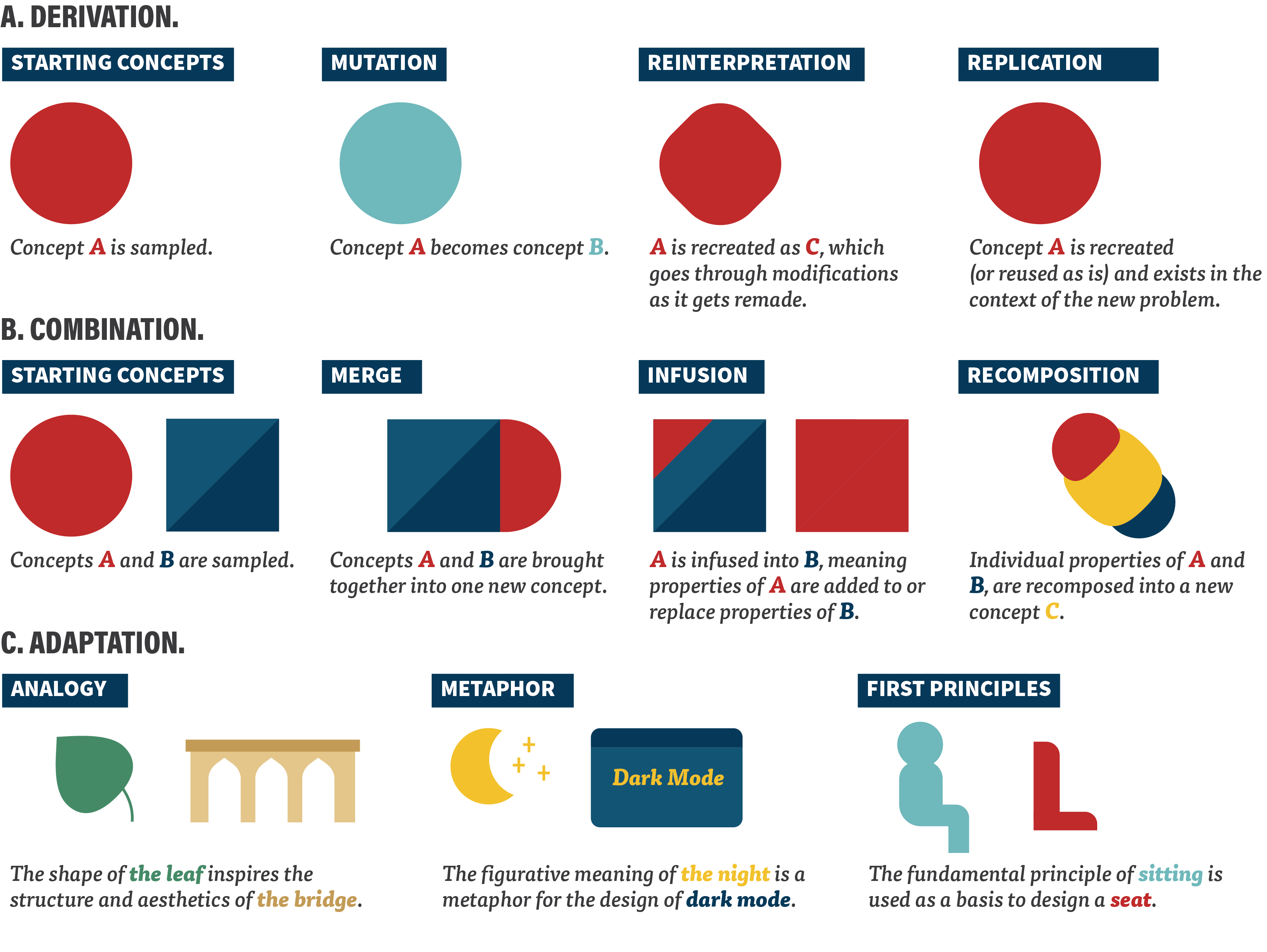}
     \caption{Procedural actions within Protosampling, these include (a) derivations, in which concept properties are changed; (b) combinations, in which two concepts are brought together; and (c) adaptation, in which concepts are transferred from other guiding principles.}
    \label{fig:procedural}
    \Description{}
\end{figure*}

\subsection{Procedural Action}
To enact Protosampling, it is necessary to operationalize the connection between the \textit{sense-making} from collected information to the \textit{action} from externalizing. Fortunately, there are theories and studies that have abstracted many of these principles. Mental representations are referred to as concepts, and they have properties with associated values \cite{smith1984conceptual, boden2004creative} (e.g., a ball has a size, colour, etc.). We outline and categorize different ways to manifest ideas, which provides a lens to understand Protosampling (Figure \ref{fig:procedural}).

\subsubsection{Derivation}
Gero  \cite{gero1994computational} as well as Cross \cite{cross1997descriptive} outline incremental changes an idea can undergo. We also account for how ideas might integrate into a creative problem \cite{sternberg2003development, makel2014creativity}. 
It is worth noting that these categories for procedural actions in Protosampling are not mutually exclusive. 

\paragraph{\textbf{Mutation.}} An incremental change that transforms a concept. This can involve one or more properties that are altered. For example, an object is brought in and its colours are changed.

\paragraph{\textbf{Reinterpretation.}} A recreation of a concept, which in the process of recreation gets slightly modified. Specific properties may change in this process. One example would be an animation including a redrawn existing painting to match the style. 

\paragraph{\textbf{Replication.}} Either a faithful recreation or reuse of an existing asset. Most properties are likely intact. For instance, assets from one project are brought in and reused in a new production.

\subsubsection{Combination}
One way in which ideas progress in creative processes is through combination. The literature uses \textit{conceptual combination} as "a mental act by which imagination brings concepts together to produce new ideas in creative processes" \cite{horng2021behavioural}. Creative theories often focus on language, and have derived six types of combination \cite{sawyer2024explaining}. For simplicity and to make combination more actionable as a concept, we distill 3 types of combination to expand on existing takes \cite{gero1994computational, cross1997descriptive}:

\paragraph{\textbf{Merge.}} Two sampled concepts are brought together into the context of the creative problem. The two concepts co-exist together. For example, images are brought together in preparation for a newspaper article.

\paragraph{\textbf{Infusion.}} Given two concepts, the properties of one concept are transferred to another concept, either changing its properties and/or adding new ones.  For example, the colour palette of an image is used to set the colour for the typography on a page.

\paragraph{\textbf{Recomposition}.} Two concepts are taken apart and the properties are used in the creation of a new concept. For instance, different samples in music can be extracted from two different songs and reused in a new composition. Note how recomposition implies a deconstruction of the source material for reassembly.

\subsubsection{Adaptation}
Creative work often benefits from bringing concepts from one domain into another \cite{sawyer2024explaining}. What distinguishes adaptation from derivation is the source and how it is used within the context of the creative problem.

\paragraph{\textbf{Analogy.}} It is a structure comparison transferring functions, logic or relationships into a domain. Christie and Gentner describe analogies as identifying patterns or threads across ideas \cite{christie2014language}. An example is a subway map, which translates physical locations to only show relevant information in an abstracted way.

\paragraph{\textbf{Metaphor.}} It is a figurative framing that transfers meaning from one domain to another \cite{sawyer2024explaining}. Unlike a combination, a metaphor relies on a concept that is not directly linked to the topic at hand \cite{ortony1979beyond}. An example of metaphors is any instance of skeuomorphism in HCI/UX design, such as folders and the file system. 

\paragraph{\textbf{First Principles.}} First principles are basic truths and assumptions about a situation that cannot be reduced further, a return to the fundamentals \cite{cross1997descriptive}. For example, a first principle when thinking of a chair is that it supports a body at rest, as opposed to being a four-legged object.  

\subsection{Situatedness}
Protosampling can be thought of as a situated activity that takes place in a workspace (whether physical or virtual). In a traditional sense, sampling is typically associated with moodboards, which enable paradoxing, aligning, abstracting and directing designs \cite{luceroFramingAligningParadoxing2012}. While creativity theory emphasizes the ideation and thinking part of creativity, and design theory describes prototyping as a practice, these domains do not often discuss what happens with the work itself. Fortunately, we can also draw further understanding on the creative process from past studies on knowledge work, as creative practitioners are a significant subset of knowledge workers \cite{bondarenko2005documents}. 

Knowledge workers process information to generate new knowledge \cite{kidd1994marks}. Like the creative process, knowledge work is described as erratic and non-deterministic \cite{bondarenko2010requirements}, with individuals working with many materials at a given time. Practitioners activities can range from highly unstructured tasks that emphasize information to more structured ones that emphasize documentation \cite{bondarenko2005documents}. In fact, the key tasks to create knowledge as articulated by Oren \cite{oren2006overview} relate directly with the creative process, as it is an interplay of capturing, organizing, formalizing, and retrieving. Materials are constantly being used and remixed \cite{henderson2009empirical}.

Knowledge work studies also emphasize the importance of engaging with the workspace -- depending on the task, information is organized (e.g., into folders) or grouped in piles \cite{bondarenko2005documents, malone1983people}, with the workspace actively providing context to help people get to the right mind frame, and the layout acting as a living reminder of the current activity. Materials are distinguished by their appearance, size and position, its proximity to tools, and overall placement all implicitly embodying clues about the task to help keep it uninterrupted. For example, a printed document can be annotated and placed on top of a keyboard with a pen on top. When the practitioner walks away, this document acts as a \textit{trail} that holds information on the current state: one can see what the document is, how many pages there are, what the current progress on the annotations is, and the placement by the keyboard indicates it is currently active. The manipulability of the physical document, flipping through pages, brings context at no cost \cite{bondarenko2010requirements}, and the need to apply mechanisms such as search decrease when a document is \textit{`hot'} \cite{sellen2003myth}. This is why Logan and Smithers argue that materials and prototypes serve to evoke memory and this helps progress moving towards a solution. Meanwhile, information that is less relevant to the current task is stored to reduce clutter \cite{sellen2003myth,henderson2009empirical}, which reduces its visibility -- it becomes searchable yet easily lost \cite{henderson2009empirical}, retrieved only for specific needs \cite{jones2010keeping} in a given context \cite{oren_overview_2006}. Seeing how information is managed sheds light to the value of creative practitioners to structure their informational search and keep relevant information at hand \cite{mobley1992process}.

It is no surprise then that creative practitioners have adopted workarounds to emulate physical spaces when working with digital tools. Frich et al. \cite{frich2019strategies} describe the use of margins around digital workspaces to keep relevant materials and past versions. This suggests an active use of Protosampling across domains. This is why there is often a desire to create tools that enable accessing and manipulating information \textit{at the right place and the right time} \cite{surfacefleet2020brudy}.

With knowledge work providing context about the importance of information within the process, it is possible to appreciate the challenged posed by the fragmentation of information. Being able to follow the \textbf{\textit{trails}} of the process might hold an additional value often missed with the active digitization of materials. Such concern becomes especially true when thinking about \textit{incubation}, in which insights are derived when stepping away from the work \cite{sawyer2024explaining, cross2011understanding}.

\section{Related Work}
This work builds upon prior systems that support sampling for ideation and generative AI interfaces that provide a variety of interactions between inputs and outputs.

\subsection{Materials as Inspiration and Samples}


Digital and physical mood boards allow creators to explore design spaces, visualize their ideas, and share their visions with others through collecting reference material arranged in a spatial manner \cite{kochImageSense}. These serve to structure the problem and align on aesthetic or conceptual directions \cite{lucero2012moodboards}, and become more effective when treated as active artifacts that evolve with the project \cite{moodboardsdesign2004mcdonagh}. Recent work frames sampling as a core creative activity where creators capture and collect materials, and even organize and remix them. Moodcubes \cite{ moodcubesIvanov} and VRicolage \cite{stemasov2023immersivesampling} are examples of systems that move beyond flat mood boards, leveraging spatial and multi-modal sampling to encourage recombination of references. 


Generative AI allows rapid directed generation of inspiration material \cite{peng2024designprompt, gancollage2023wan} and supports quick ideation in a variety of visual domains \cite{aideation2025wang, dalle, visualstories2024antony}. GanCollage \cite{gancollage2023wan} integrates generative AI into mood boards, tagging generated images \revision{with semantic labels used to explore the latent visual space. These interactions focus on exploring variations rather than on composing or editing artifacts.} Other systems augment canvas mood boards with semantic tagging and search as a way to better navigate and reflect on these spaces that might have less visual organizational structure \cite{peng2024designprompt, kochImageSense, kochSemanticCollageEnrichingDigital2020}. These works treat found and generated artifacts as references to be used later in the creative process, where the making is executed.

\subsection{Creating Media with Generative AI Interfaces}
Generative AI systems vary in how they treat the relationship between inputs and outputs, how generations are produced, and the arrangement of these interactions. 

\paragraph{\textbf{Prompt-Based Interfaces.}} Systems such as ChatGPT \cite{chatgpt} and Midjourney \cite{midjourney} primarily use natural language prompting as an accessible method to generate output with AI. \revision{These interfaces present a linear thread of prompts and outputs where one thread of exploration is active at a time}. The outputs are inherently arranged chronologically, burying previous generations in history. These approaches are easy to use, but they lack controllability of outputs or discoverability of possible operations.

\paragraph{\textbf{Node-Based Interfaces.}} On the other end of the spectrum of control are systems like ComfyUI \cite{ComfyUI} that expose the models' internal parameters through a node-based interface and allows users to chain modules together to create workflows for generating media. These offer a higher degree of control over the final generation, with the focus still on executing a single workflow at a time. The final output can then be piped back as the input for a new workflow, where the intermediate steps are erased. \revision{Node-based interfaces prioritize visible components for \textit{atomic operations}, where each node can load files, run operations, or save outputs.}

\revision{Some systems look to bring node-based programming to a more approachable level, such as FLORA \cite{FLORA}, Runway workflows \cite{RunwayML}, and Figma Weave \cite{FigmaWeave}. These tools turn media into nodes, enabling use of multiple images and revealing outputs as they are processed within an explicit graph structure. These systems emphasize the \textit{links} between objects, highlighting the \textit{operations} taking place. The thinking processes still primarily focus on graph-based workflows, where outputs are cleared and re-generated every time the workflow is executed. In fact, FLORA describes its own approach as \textit{"built for professionals who think in systems, not just outputs"} \cite{FLORA}.} 

\paragraph{\textbf{Bringing Direct Manipulation to AI Generation.}} Between these two extremes of emphasizing prompts and emphasizing operations, some are systems provide direct manipulation and combination text prompts with other modalities, such as sketching \cite{promptpaint2023chung, sketchflex2025lin, scaffoldingsketch2024, designsketches2023zhang}, image references \cite{leong2025paratrouper, realfill2024tang}, region masking \cite{dang2023worldsmith}, composable prompt widgets \cite{composable2025amin}, and other methods for refining prompts and iterating through outputs. 

\subsection{Virtual Canvases}
\revision{Canvas interfaces are 2D planes where content can be imported, created, or modified, some made for brainstorming and collaboration, such as Mural \cite{Mural} or Miro \cite{Miro}, whereas others are more optimized to act as a board for note-taking, such as MilaNote \cite{Milanote}. Canvases enable arranging media in space, keeping materials simultaneously visible and supporting branching, nonlinear exploration \cite{kochImageSense}. Where other work has explored attributes (such as shape or color) as first-class objects that can be manipulated \cite{xia2016objectorienteddrawing}, support free-form drawing and collecting resources \cite{hinckleyInkseine} or coupling media with tools as the focus \cite{surfacefleet2020brudy, walny2016thinking, xia2016objectorienteddrawing}. }

\revision{With generative AI, new interfaces for canvases have started to emerge. ImaginationVellum \cite{marquardt2025imaginationvellum} uses 'generative strokes' to rapidly create variations of different design ideas leveraging spatial relationships to text (e.g., labeling styles) modulate between concepts. Adobe Firefly Boards \cite{Adobefirefly}provide canvases for organizing and displaying generated content. Firefly Boards provides a global generalized interface more similar to a prompt-based text box at the bottom centre of the screen with the option to import style and structure references, offering access to many online models.}

\subsection{Towards Operationalizing Protosampling}
\revision{Generative AI has begun to reshape creative workflows across visual domains \cite{misty, vrcopilot, neuralcanvas, creativeconnect}, accelerating exploration and enabling rapid generation of artifacts. The speed and capabilities of generative AI have evolved to increase the collapse of sampling and prototyping, merging thinking and making into a more closely coupled process. Systems such as Firefly Boards \cite{Adobefirefly}, Paratrouper \cite{leong2025paratrouper} and ImaginationVellum \cite{marquardt2025imaginationvellum} are already leaning towards that convergence. Firefly Boards provides a variety of online models and its image generation can use one style image and one structure reference. Paratrouper \cite{leong2025paratrouper} allows means for combination and derivation while providing more structured exploration and hiding away reference material. ImaginationVellum \cite{marquardt2025imaginationvellum} excels at derivation and sketch integration into images. However, these systems emphasize particular aspects rather than fully integrating sampling and prototyping as a unified spatial practice. }

\revision{
With Atelier, we operationalize Protosampling through a media-first canvas providing access to a variety of local open-source models and algorithms to manipulate and generate content. Interaction is presented in modular widgets abstracting activities (Draw, Paint, Trace, Modify, Animate) so that creators can focus on the task rather than the model selection or workflow authoring as done in tools like ComfyUI \cite{ComfyUI}. At the same time, these widgets show every available input that can be used, and materials - both inputs and outputs co-exist in the same visual space. Thus, we restrict the high flexibility of a node-based paradigm while making a wide variety of parameters to be accessed and tweaked. We also translate terms in ways that can be useful for creators without using overly technical terms (e.g., renaming \textit{"denoise percentage"} to \textit{"preserve original image"}), and hide away features that require more technical expertise, such as selecting a sampler and a scheduler. Thus, its novelty lies on the approach, the interactions, and the abstractions.}

\section{Design Rationale}

To develop a system enacting Protosampling for visual media, we combined (1) learnings from prior work, (2) qualitative observations of online videos of creators working with image and video generation, (3) creativity theory and (4) our own experience testing the system as we continuously prototyped and iterated on it \cite{ledo2018evaluation, neustaedter2012autobiographical, stappers2017research, zimmerman2007research}. These shaped our design rationale to address the challenges and opportunities for supporting thinking and making through Protosampling.


\paragraph{\textbf{R1. Blending Spaces for Thinking and Making.}} Tightening the loop between sampling and prototyping requires a workspace that invites open-ended exploration and creation while also bringing materials together. Materials should be able to be brought in, generated, and recombined. For this reason, an interface like this should be media-first and take advantage of spatial arrangements for meaning and organization, thus requiring a canvas-like interaction. This also supports hot and cold areas \cite{sellen2003myth}, where creators can pile information according to their needs. To invite equal usage of sampled and generated media, they should be indistinguishable.


\paragraph{\textbf{R2. Encapsulating Technical Workflows by Activity while Offering Control.}} Working with multiple types of generative AI models means that creators are typically required to think of the model they need to use and determine the associated settings and parameters. This process can be quite technical and break the creators process. It is important to determine which parameters should be exposed so that creators have agency and control, but do not need to spend significant time learning or tweaking parameters. The interface must expose the right level of detail for the models, and use parameters that generate consistent and coherent media.

\paragraph{\textbf{R3. Highlighting the Process through Trails.}}
A natural result of working with non-linear and freeform canvas is that there is no longer chronological order or direct input-to-output visual connection to the items. For this reason, it is important to have a rich provenance history for each asset and to expose this in a way that allows creators to easily revisit assets and make sense of how they came to be, what inputs created it and what came downstream of it, and be able to see these connections between items. These should allow creators to retrace their steps and make sense of past decisions, as well as revisit earlier points in their process to explore new alternative paths.

\paragraph{\textbf{R4. Supporting Organization, Collections and Explorability.}}
Because a canvas for thinking and making represents an instantiation of the creative process, it is important for creators to be able to organize and explore information. Creators must have a means to access materials at the right place and the right time. While part of this should be afforded by the spatial arrangements creators may naturally use, they should also be able to search through materials and create relevant collections to easily access important items. 


\section{Atelier}

To operationalize \textit{Protosampling}, we designed Atelier, (Figure \ref{fig:teaser}) where thinking, making, and reflection happen in the same space. Atelier treats the canvas as a living workspace where media are first-class objects that can be arranged freely to be decomposed, sampled, generated, remixed and recombined. 


\subsection{Overview}
The canvas acts as the workspace akin to an artist's studio where materials are brought in, arranged, and manipulated to create new output. We support drag-and-drop upload of images, video, text, audio, and 3D models, as well creating text and sketches directly in-canvas. All media assets can be moved and grouped, then immediately transformed through quick operations that decompose them into reusable components, allowing experimentation and manipulation of materials before moving to more complex processes. 

Because the canvas affords spatial layouts, creators can organize information freely, ad-hoc arrangements can be made that develop meaning over time. Having a mix of collected and generated assets can stimulate thinking with on-going engagement around potential triggers for reinterpretation \cite{eckert2000sources}.

In physical spaces, interactions often are intimately connected with the setting where they occur \cite{dourish2001action}. This is especially obvious in an artist studio, where distinct 'places' hold meaning of an activity, nearby materials are connected to the task at hand, and materials might migrate over time or cycle back. This inspired our design to localize activity:\textit{quick operations} can be made directly on media, and larger operations can be made in localized workstations, which we call \textit{Easels}, to provide nuanced controls and structure. Any creation is brought directly into the canvas and treated as potential new material that co-exists with the rest.

Past work often discusses how applications bind work to a given time (e.g., \cite{surfacefleet2020brudy}). The state of the workspace thus reflects the lead-up from the process up to that point, its history holding important information. This motivated us to explore means for \textit{organization, sense-making} and \textit{provenance}.

\begin{figure}
    \centering
    \includegraphics[width=\columnwidth]{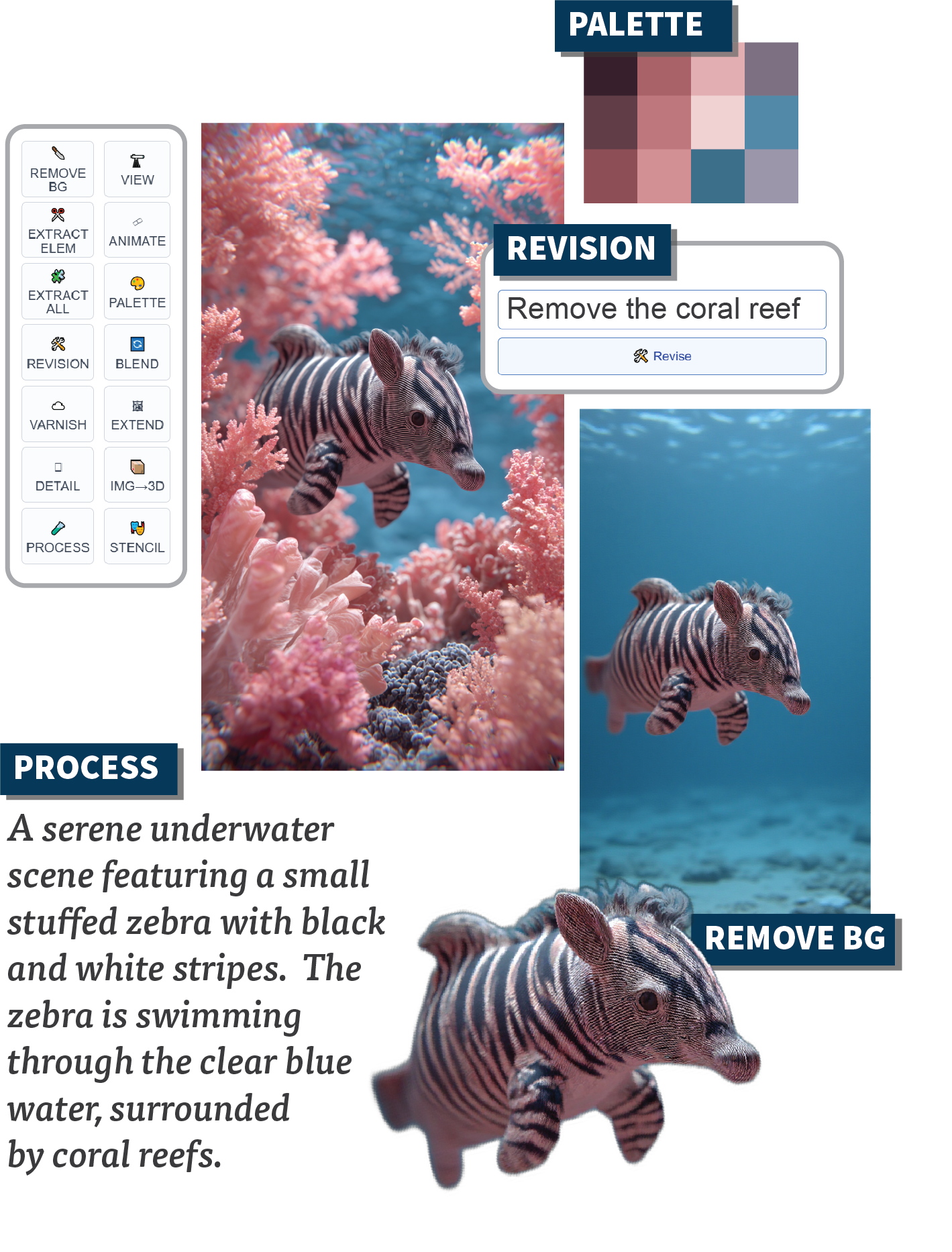}
     \caption{Quick operations. Images in Atelier present common functions for image processing. Illustrated examples include: revision, which allows making edits to the image; process, which analyzes the image, generates ControlNet preprocessors and captions it; and remove background. }
    \label{fig:quickInst}
    \Description{Quick operations. Images in Atelier present common functions for image processing. Illustrated examples include: revision, which allows making edits to the image; process, which analyzes the image, generates ControlNet preprocessors and captions it; and remove background.}
\end{figure}

\subsection{Quick Operations}
Immediately available actions on media are those that require little to no additional input, such as generating \textbf{Quick Sketch} on text, using the contents as a prompt to generate an image. With images, creators can quickly decompose the material into more reusable components (Figure \ref{fig:quickInst}) with \textbf{\textit{Remove Background}} (retains foreground items on a transparent background), \textbf{\textit{Extract Element}} (extracts targeted elements such as "the blue flower"), \textbf{\textit{Palette}} (creates a color palette), and \textbf{\textit{Stencil}} (producing structural control images such as depth mask, line art, poses, etc that can be used to control later generations). Creators can also refine images and make small adjustments with \textbf{\textit{Revision}} (single minor revision such as \textit{"give the bunny a mustache"}), \textbf{\textit{Upscale}} (upscales the image to a higher resolution and adds detail), and \textbf{\textit{Blend}} (makes style and lighting consistent). Lastly, they can create extensions of the image with \textbf{\textit{Extend}} (growing the image beyond the current frame), and \textbf{\textit{View}} (generates different perspective views), \textbf{\textit{Quick Animate}} (similar to Quick Sketch, generates a 5 second video from the video without additional prompting), and \textbf{\textit{Sculpt}} (converts the image to a 3D model). 3D models can be manipulated and rotated and captured as new images that then serve as character and perspective references.

\subsection{Easels: Spatial Modular Workstations} 

Easels are stations where the workspace's materials are gathered and use to generate new media. As a parallel to studio workflows where materials are physically gathered around an area to prepare the act of using them, within Atelier, assets are moved in proximity to an easel before their use. Each easel encapsulates an act of generation with different goals and thus exposes different controls, which are abstractions of distinct ComfyUI workflows in the backend. These generations are non-destructive: inputs are preserved, can be reused, and iterated on. Easels aim to function as flags for provenance, where the presence of an easel and its surrounding materials signals a complex creative intent in the area. 

\subsection{Easels to Prepare Material}
We offer two lightweight easels to help ideating and drafting concepts before moving to more complex generations. They provide spaces for gathering, arranging, and externalizing ideas, similar to creating thumbnail sketches before moving onto the final piece. 

\begin{figure}
    \centering
    \includegraphics[width=\columnwidth]{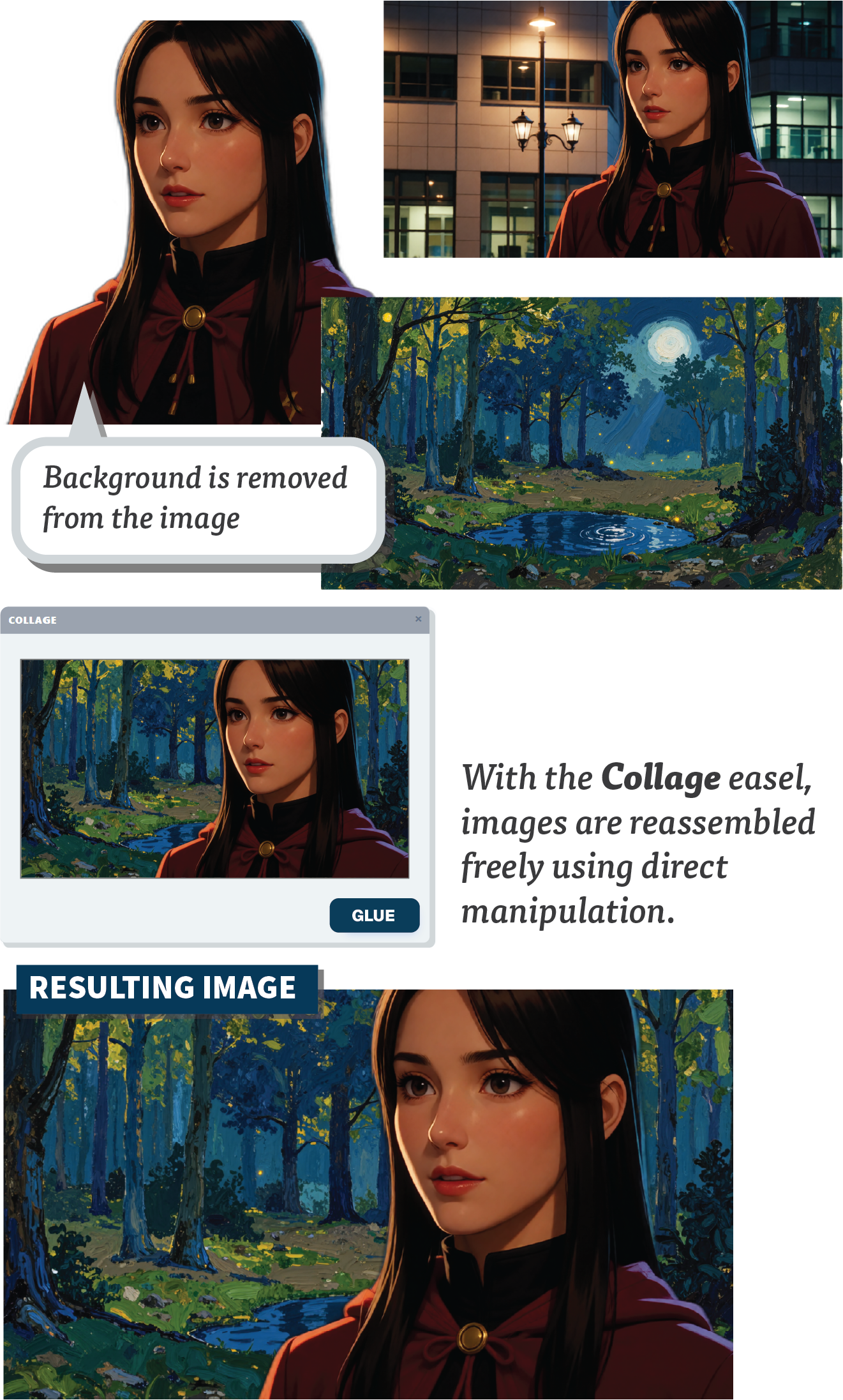}
     \caption{The Collage easel allows composing images into a new image. This example one character extracted from a background and brought into a new scene.}
    \label{fig:collageFig}
    \Description{}
\end{figure}

The \textbf{Collage} easel (Figure \ref{fig:collageFig}) is an area for composing multiple images in a freeform arrangement and then "gluing" those together to create a new composite image that can be reused. This can be used to block out character and object arrangements in a scene. For example, a creator might extract a character from one image, use another generation as the background, and collage them together while adjusting size and position. The \textbf{Sketch} easel allows freehand drawing that can guide future image or video generation. Having these prep easels that output images allows for reuse of sketches and collages over time and invites iterative adjustments.

\subsection{Easels to Generate Material}
Image easels (Draw, Paint, Trace, Modify) contain the workflows for generating new images with abstracted fine levels of control blending text prompts, image references, parameter sliders, and image masks. The Animate easel enables generating videos from images acting as start or end frames. 


\paragraph{\textbf{Draw.}} The Draw easel (text to image) takes in text as the \textit{positive prompt} of what the creator wants. Optional input(s) include: preset \textit{style options} (Realism, Dreamlight, Anime, Retro Anime, Animated, 3D, Pixel Art), a \textit{details} slider to set the amount of detail in the final image, an \textit{adherence} slider to set how strongly the model adheres the prompt, a \textit{start image}, and a \textit{preserve} slider that sets how much percentage of the original image is preserved. The use of a \textit{start image} is akin to bringing in an already existing painting and painting on top of it, where original features may show through, but ultimately a new painting is generated \cite{ledo2025generative}. 

\begin{figure*}
    \centering
    \includegraphics[width = 1.0\textwidth]{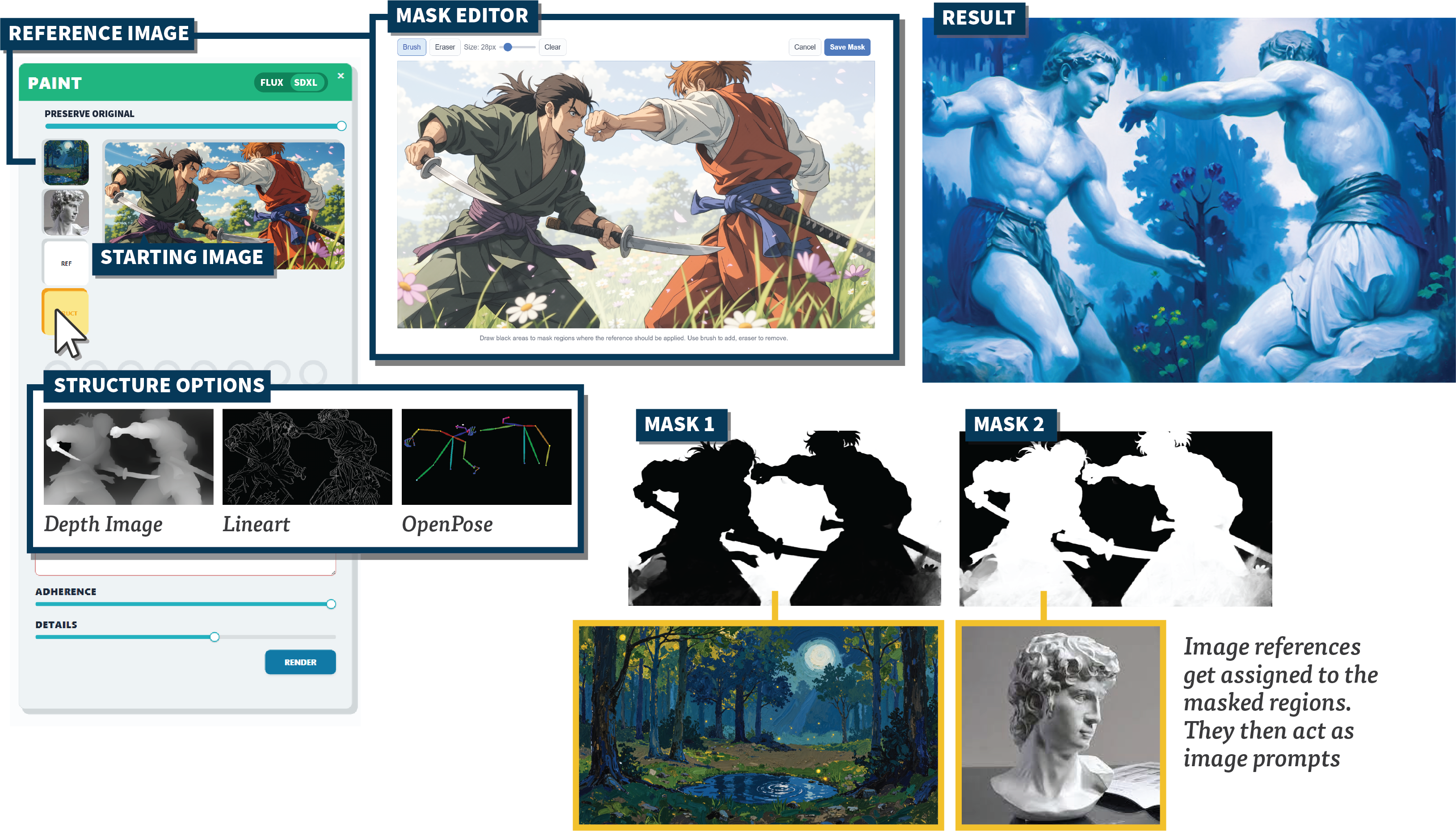}
     \caption{Description of the Paint easel showing how one might add reference images, a starting image and a structure image. The reference image box allows setting the strength and also opening a mask editor to determine the area of influence for that particular reference.}
    \label{fig:paint-easel}
    \Description{}
\end{figure*}

\paragraph{\textbf{Paint.}} The Paint easel (Figure \ref{fig:paint-easel}) extends the capabilities of \textit{Draw} by enabling up to 3 \textit{image references} on top of text-to-image generation with adjustable strengths of influence. Additionally \textit{reference masks} can be drawn to target areas where those references are applied. A \textit{structure image} can also be included, chosen from any image on the canvas. These \textit{structure images} are controls for the layout (depth, line, scribble pose) with adjustable strength. An additional prompt box enables \textit{negative prompts}, where the creator can specify what they do not want in addition to the \textit{positive prompt}. The \textit{style options}, \textit{details}, \textit{adherence}, and \textit{preserve} sliders are the same as in \textit{Draw}. 


\begin{figure}
    \centering
    \includegraphics[width=1\columnwidth]{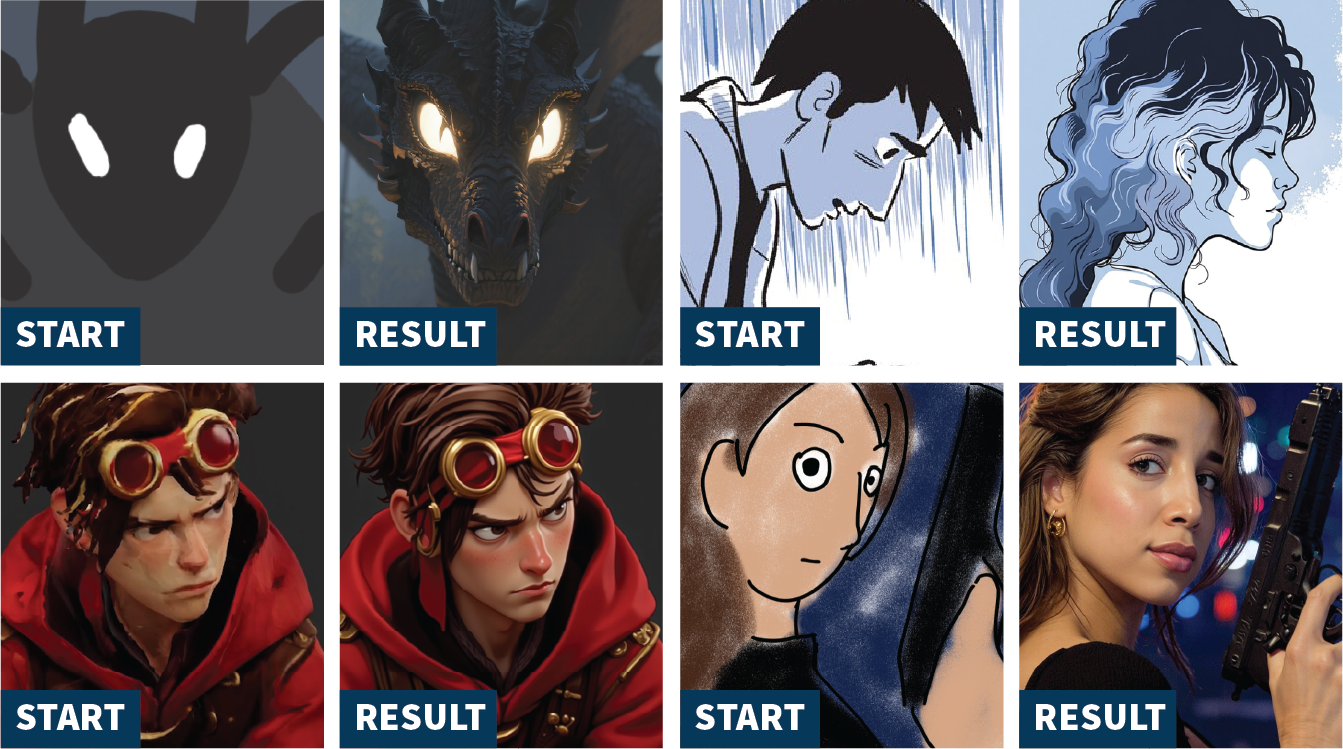}
     \caption{Examples images created with FlowEdit (Trace). }
    \label{fig:trace}
    \Description{
    }
\end{figure}

\paragraph{\textbf{Trace.}} The Trace easel supports generating a new image based on an existing image, selectively redrawing their content and structure. By using two prompts, one to describe the current image and one to describe the desired image, the system can deconstruct an \textit{input image}. Similar to the Draw and Trace, sliders for \textit{prompt adherence}, \textit{style options}, \textit{details} and \textit{preserving the original image} are present. An additional range slider for the \textit{retracing range} determines when in the deconstruction to redraw. The easel includes optionally a \textit{structure image} from the input, \textit{style options}, \textit{details}, \textit{preserve} and \textit{adherence} sliders. The trace operation works especially well with coloured drawings, rough 3D models, or for image restyling (Figure \ref{fig:trace}).

\paragraph{\textbf{Modify.}} The Modify easel allows for a broad range of edits to be applied. To support further discoverability, we extracted a set of cinematographic edits. One can change the aspect ratio of the image without major distortions. Atelier provides a set of \textit{relighting} presets for color temperature, direction of light, and light intensity, as well as a set of camera angles, and styles. These suggestions, when clicked become additive \textit{prompt pills}, which are preset text strings that get appended to the prompt, and creators can additionally specify their own custom changes in the prompt. Similar to the other draw easels, there is the option for  \textit{adherence} and \textit{detail} sliders. 



\paragraph{\textbf{Animate.}} The Animate easel creates a video by offering the ability to specify images for the \textit{first frame} and/or \textit{last frame}, where the output video creates a smooth transition between the two frames, guided by the \textit{positive} and \textit{negative prompts}. Both frames are optional, so where one can specify a \textit{start frame} without an \textit{end} and vice versa, or do direct text to video generation by not providing either frame. 
It is also possible to choose a camera motion from a set of predefined motions with that creators can add to their prompt through \textit{prompt pills} that additionally serve as examples of how creators can structure their own motion prompts. The resulting video is added directly to the canvas, where it sits alongside the other assets and can be decomposed into still frames. 





\begin{figure*}
    \centering
    \includegraphics[width = 1.0\textwidth]{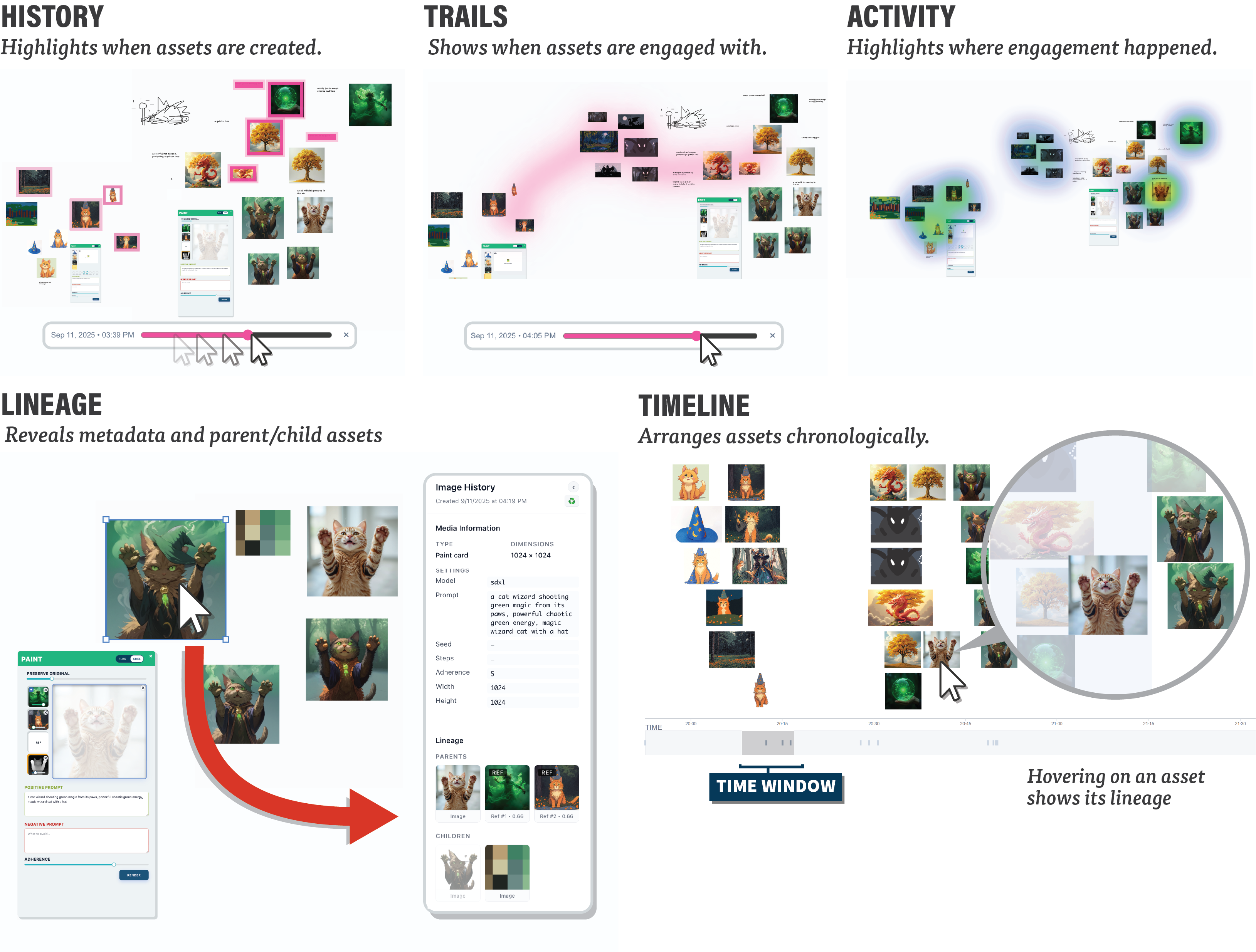}
     \caption{Examples of Provenance in Atelier. Figure shows history, trails, activity, lineage and timeline.}
    \label{fig:provenance}
    \Description{}
\end{figure*}

\subsection{Provenance, Organization and Sensemaking}
To give structure to a nonlinear and non-chronological freeform canvas, we provide features that support provenance and organization woven directly into the canvas. Provenance in Atelier acts as a living trail, where each asset carries its lineage information such as whether it was imported or generated, the type of easel that generated it, and the parameters and assets used (Figure \ref{fig:provenance}).

\paragraph{\textbf{Lineage.}} Whenever an asset is selected, the lineage panel shows all of the relevant information, including the parent assets that generated it and all the children assets that came from it, highlighting those in the canvas and allowing for quick navigation to each. Deleted images are not present in the canvas, but still remain recorded in the provenance graph and can be brought back as an active asset. Each asset that was generated by an easel carries a \textit{recreate} button that allows creators to show or re-instantiate the easels that generated an image or video with the exact same parameters and continue iterating from there. \revision{From a given image, using the lineage panel, one can trace back all of the references and parameters that generated it by recursively looking at the parent. At any point in time, they can choose to recreate the easel, which would allow them to work non-destructively from a branching point in the past. }

\paragraph{\textbf{Asset Emphasis.}}
Outputs might have varied importance, as some might be more useful than others, or closer to the creator's intent. Thus, we added an option to de-emphasize assets, which reduces their opacity. 

\paragraph{\textbf{History.}} Examining history features a slider that highlights media on the canvas in chronological order with timestamps, showing the progression of creation across time and space with a sliding window of 5 elements at a time. 

\paragraph{\textbf{Trails.}} Another way to represent interaction over time is through \textit{trails}, which shows a diffused trajectory that highlights where in the canvas engagement took place at a given time, aggregating the information shown in history, summarizing the overall path rather than individual elements. 

\paragraph{\textbf{Activity.}} We can also show engagement in a more absolute fashion through a \textit{heatmap} overlay. We keep track of the time of creation of an asset, the last time it was selected, and the total number of times an asset was clicked or used as a proxy for how much time was spent with it. This overlay uses click count to show the areas of activity, emphasizing the areas and assets creators have spent the most time. 

\paragraph{\textbf{Timeline.}} The overall activity over time can be seen in a timeline, which spatially arranges assets on a visual timeline, where hovering on an asset highlights the parent and the children.

\paragraph{\textbf{Organization.}} Every image is automatically tagged with an AI generated caption. This allows for a canvas wide search through the caption, prompt and parameter texts to find media through keywords. The canvas also supports grouping assets and automatically arranging them into grids. This helps appreciate many materials in close proximity, and also chunking them into a single selection. To support longer term cold storage, we support \textbf{Collections}. The recurring characters, scenes and other groups of media can be saved into collections that can be tagged. Then at any other location in the canvas, the creator can pull a copy from these collections to quickly reuse their saved media. These collections prevail throughout the whole project, even if one is working on a different canvas page. The images and videos that creators want to save can be added to the \textbf{Exhibit} gallery, captioned and rearranged, whether these are frames in a storyboard for a film or final assets for a standalone project. This serves as a space to keep important assets that are specific to the current project. These capabilities support reflection of the creation process, tracking connections between media beyond history, but as a way to navigate the canvas and find relevant information.





\section{Usage Scenario}
To illustrate how some of the features of Atelier come together, we present a use case scenario for Atelier. Fiona is a filmmaker who wants to ideate a short fantasy film concept about a warrior that kills a dragon.

\begin{figure}[!ht]
    \centering
    \includegraphics[width=1\linewidth]{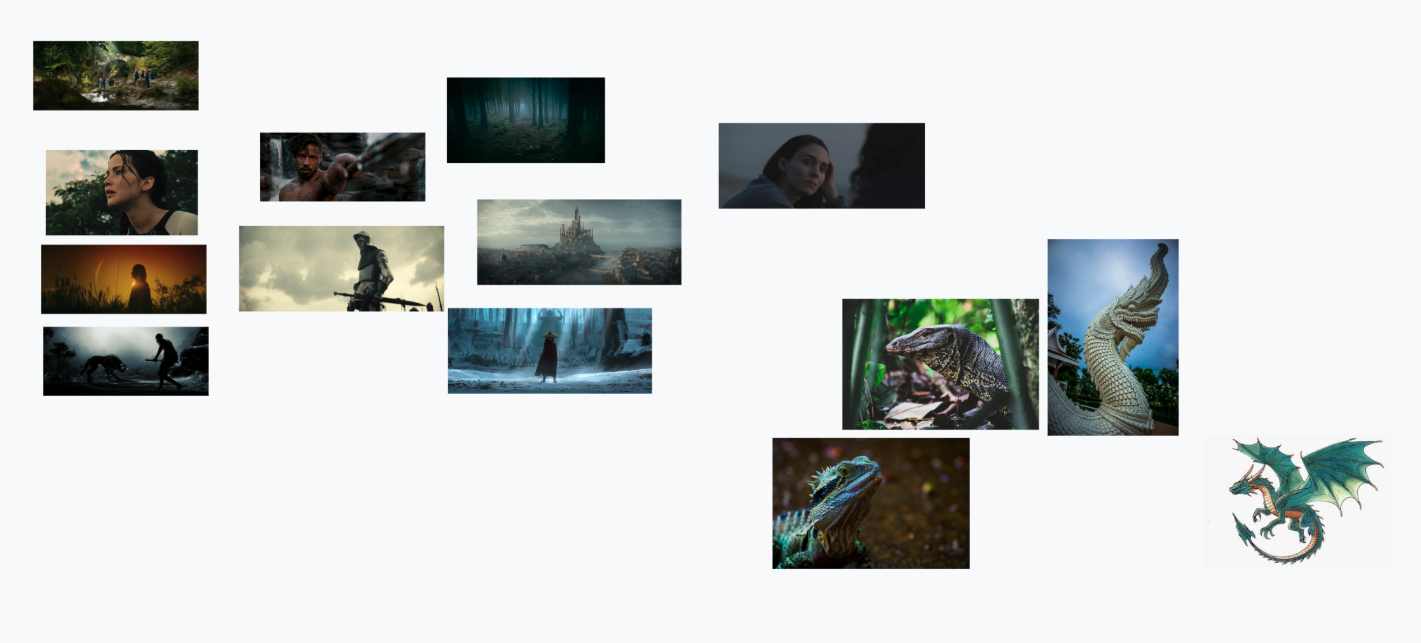}
    \caption{Initial canvas with drag-and-dropped inspiration images.}
    \label{fig:moodboard}
\end{figure}

Fiona opens Atelier and import images of dragons and lizards, epic hero shots, and fantasy scenes that carry the aesthetic of world they want to create (Figure \ref{fig:moodboard}). Looking at these images arranged in the canvas, acting as a mood board, they get a sense of the atmosphere of the story.

Fiona wants to settle on a protagonist first. She writes out a few descriptions and uses \textbf{Quick Sketch} to rapidly render images and likes a concept for the prompt "female warrior wearing a cape" but dislikes how the character is facing directly forward. She uses \textbf{Remove Background}, then \textbf{Sculpt} to turn her into a rough 3D model and is able to adjust her warrior to any angle and get still captures of different views.

\begin{figure}[!ht]
    \centering
    \includegraphics[width=1\linewidth]{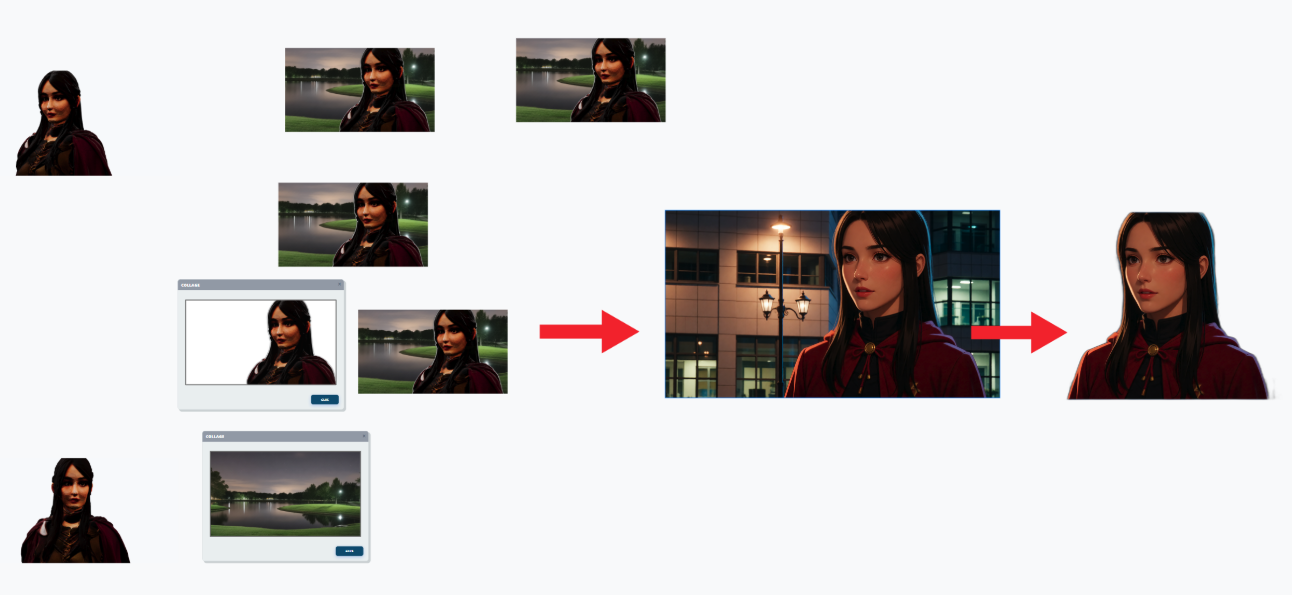}
    \caption{The 3D model is used to create different views of warrior.}
    \label{fig:character1}
\end{figure}

She wants to see her warrior in a scene, so she opens a \textbf{Collage} easel and arranges an image of her warrior on a scene of a lake from her inspiration images for a close up shot. Taking this glued image, she then opens a \textbf{Trace} easel, and describes her original image as "a woman wearing a red cape in front of a lake" and her target image "a woman wearing a red cape, painting style." The generated image does not have the background she wanted, but Fiona likes the newly stylized version of the warrior. She uses \textbf{Remove Background} again and then saves the image into a new collection titled "Warrior." 

\begin{figure}[!ht]
    \centering
    \includegraphics[width=1\linewidth]{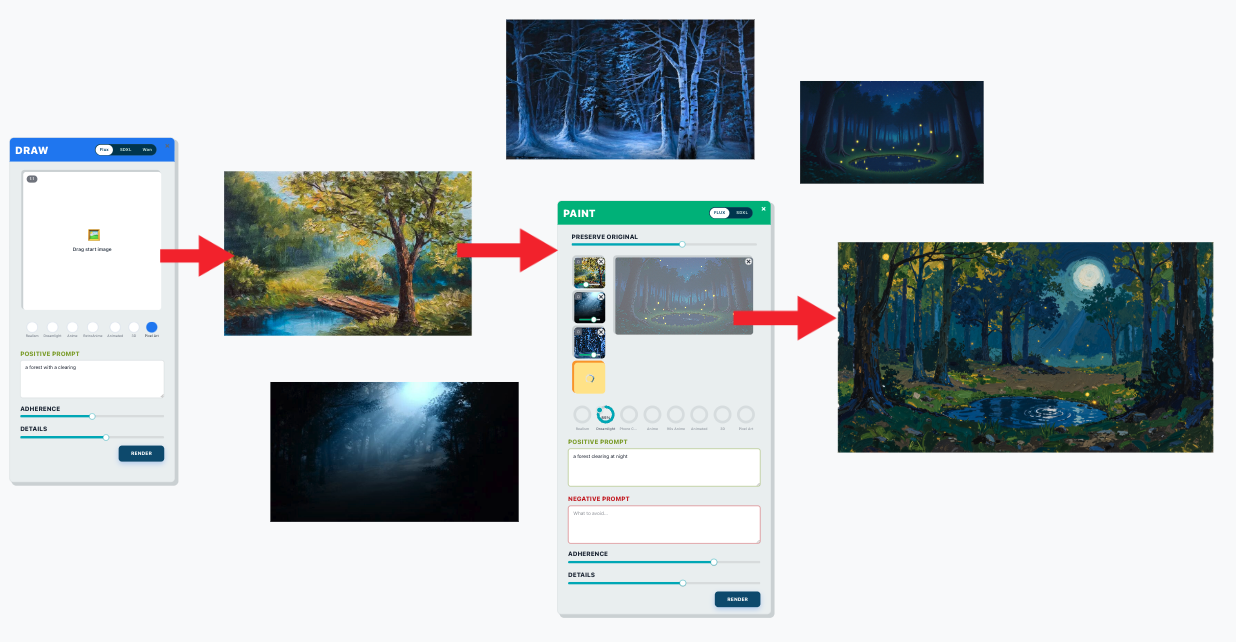}
    \caption{Designing the Setting. A forest is generated and then used as a style reference along other images of forests.}
    \label{fig:setting}
\end{figure}

She now focuses on the setting for the scene: a forest. She continues building on painting styles, using the \textbf{Draw} easel with "a forest with a clearing" (Figure \ref{fig:setting}). She likes the result, but wants it to be night time, so she uses \textbf{Revision}. One of the results surprises her, giving her new ideas, and she is drawn to the spooky feel and decides to move in that direction instead. She opens a \textbf{Paint} easel and references the spooky forest and other images with the prompt "a forest clearing at night". She likes the paint style of the first generation and includes it as a style for subsequent generations. 
\begin{figure}[!ht]
    \centering
    \includegraphics[width=1\linewidth]{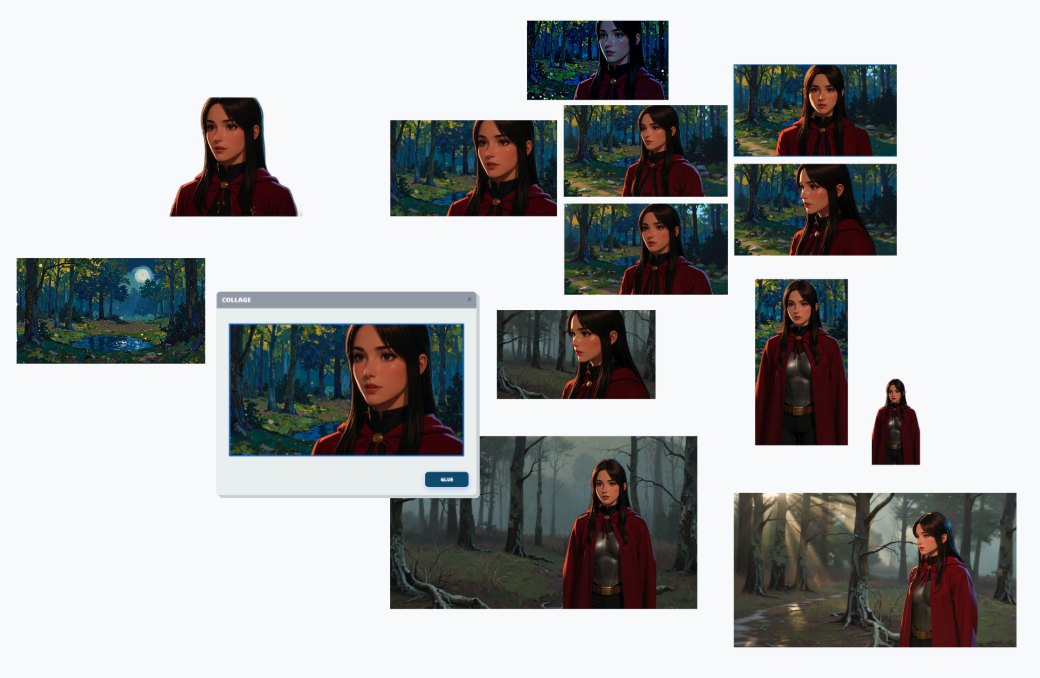}
    \caption{Exploring camera angles and staging of the warrior character}
    \label{fig:warrior}
\end{figure}
She opens the \textbf{Collage} easel and composes another close up of the warrior image taken from the collection and the forest scene. Fiona is unsure if this combination works well, so she uses \textbf{Revision} to try out different views with directions (Figure \ref{fig:warrior}) such as "make the woman face forward" and "show woman from waist up". From these, she likes some of the zoomed out views, so Fiona uses \textbf{Remove Background} again to extract the warrior in the angles she wants. Now she has a few angles and zoom levels for her scenes. Next is the dragon.
\begin{figure}[!ht]
    \centering
    \includegraphics[width=1\linewidth]{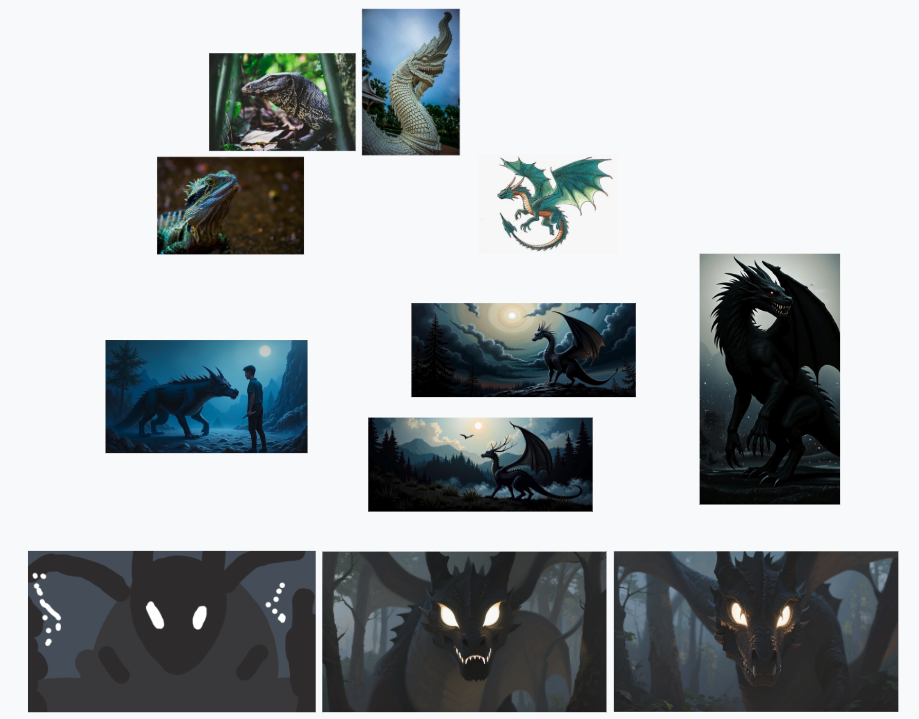}
    \caption{Character generation explorations for Dragon.}
    \label{fig:dragons}
\end{figure}
Going back to the earlier imported images of dragons and lizards, Fiona plays around with different easels and combinations to generate dragon designs (Figure \ref{fig:dragons}). She eventually lands on a colour scheme that is gray and dark. This inspires her to think of the dragon scene as one that is thematically darker than the rest. She uses these gray colors in the \textbf{Sketch} easel to draw out her dragon, and then \textbf{Trace}, this time with her drawing as the input. After iterating on the parameters, she gets a design she is happy with. She then uses the \textbf{Modify} easel to get 16:9 aspect ratio shots with a variety of different camera angles and lighting. Her favorite ones get added to a new "Dragon" collection. 

\begin{figure}[!ht]
    \centering
    \includegraphics[width=1\linewidth]{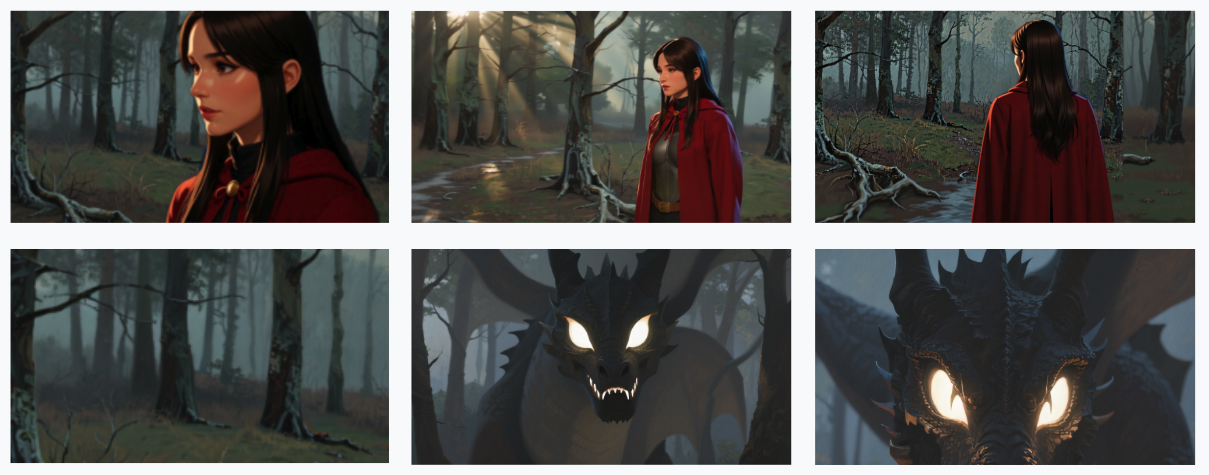}
    \caption{Final frames in the storyboard showing a female warrior in the forest as a dragon emerges.}
    \label{fig:storyboard}
\end{figure}

Observing the canvas, Fiona has an idea for a sequence: the warrior standing in the forest, the camera panning from her face around to her back, then pushing into the trees until the dragon emerges. She chooses the images of the warrior in the forest, the empty forest, the dragon approaching, and finally a close-up of the dragon, and adds them to the gallery. Reviewing the sequence, she realizes she is missing a crucial frame that positions the camera behind the warrior, and uses \textbf{Revision} to create it. From these frames (Figure \ref{fig:storyboard}), she can use the \textbf{Animate} easel to pass in these images as start and end images to create a continuous video.

\section{Implementation}
Atelier was built using tldraw \cite{tldraw} an infinite canvas SDK with built-in text and image features as well as shape management. Each item on the canvas is a shape, and shapes can point to assets (images, videos, etc.) that are hosted elsewhere. We implemented easels, generation logic, provenance, and organization features as custom React components layered on top of the canvas, using Cursor \cite{Cursor} to assist with implementation. All code was edited and reviewed by two of the authors. Each AI operation was executed by custom ComfyUI workflows (represented as JSON files).

Provenance works as a directed acyclic graph where nodes maps to an item on the canvas treated as the source of truth for canvas history. Images, videos, and 3D models may have multiple copies on the canvas, where each copy is created as a new node pointing back to the original asset. These copy nodes preserve the lineage and metadata information from the original source. Parent nodes are any media that was used to create the current child node (reference images, start images). Each node key metadata (creation time, last interaction time, number of clicks, generation parameters, and links to the parent and child nodes). Time of creation is used to create the history view and number of clicks is used to create the heatmap overlay. The timeline was built using D3\cite{bostock2011d3}. 

\subsection{Workflow Implementations: General Strategies}
We tested the Atelier ComfyUI server on computers with RTX 5090 and RTX 4090 graphics cards. These workflows use various local models, including Stable Diffusion XL \cite{stablediffusion}, FLUX-dev and Flux Kontext \cite{flux}, and Wan 2.2 \cite{wan2025wanopenadvancedlargescale}.
Three key strategies apply to many of our workflows which are part of what differentiates them from currently publicly available workflows. The first two are about how images are sampled, and the last strategy is about how to deal with multiple permutations within a workflow. \revision{An overview showing how all workflow parameters got translated into easel interfaces can be found in Table \ref{fig:rationale-table}.}

\subsubsection{Custom Samplers and Lying Sigmas}
One novel aspect offered by Atelier's workflows is the ability to add details to images. This is done through Lying Sigmas, a technique that injects noise in the diffusion process\footnote{https://github.com/muerrilla/sd-webui-detail-daemon/}. We use the Detail Daemon node\footnote{https://github.com/Jonseed/ComfyUI-Detail-Daemon}. Among the Detail Daemon pack, the Lying Sigmas sampler uses a single parameter for details (dishonesty factor). We choose values between -0.05 to 0. To support Lying Sigmas, workflows need to replace the traditional KSampler\footnote{https://blenderneko.github.io/ComfyUI-docs/Core\%20Nodes/Sampling/KSampler/} with the more complex Custom Sampler, which supports manipulating the guider and the sigmas.

\subsubsection{Normalized Attention Guidance}
Many AI models offer negative prompts. However, transformer architectures such as FLUX do not provide a negative prompt because they have 'weak guidance' (Classifier-Free Guidance value of 1.0) \cite{ho2022classifier}. A workaround is Normalized Attention Guidance \cite{chen2025normalized}, which enabled incorporating negative prompts into the system\footnote{https://github.com/ChenDarYen/ComfyUI-NAG}, which we set to a value of 9.0 (recommended values are 5.0 to 11.0). This algorithm also enables using negative prompts for models that forego higher guidance for speed-up LoRAs, such as Wan 2.2. Normalized Attention Guidance can be attached to custom samplers.

\subsubsection{Dealing with Inactive Nodes}
ComfyUI workflows are polled with every new prompt. For certain workflows, such as Paint, which take multiple inputs, there are 4 strategies one can take to execute the API: (1) programatically editing JSON files to rewire nodes depending on available inputs (e.g., Paratrouper \cite{leong2025paratrouper}); (2) creating workflows for each permutation of available inputs; (3) setting parameters to 0 so they do not affect the generation (which is not supported by some features); and lastly, our chosen approach which is to use switches\footnote{https://github.com/ltdrdata/ComfyUI-Impact-Pack}. Switches dynamically reroute nodes in the workflow by setting a single number, thereby reducing potential mistakes compared to rewiring.

\subsection{Small Workflows and Premade Workflows}
Many of the quick operations are done with ComfyUI workflows with a few nodes or reusing nodes available online, such as using template remove background, extract element (GroundingDINO\footnote{https://github.com/storyicon/comfyui\_segment\_anything}), palette, upscale, extend (outpaint) and stencil (control images). Several of these actions are pared down versions of existing easels with no extra settings (Quick Sketch and Draw; Revision, View and Modify, Quick Animate and Animate). Sculpt employs the Hunyuan3D wrapper workflow \footnote{https://github.com/kijai/ComfyUI-Hunyuan3DWrapper} to go from image to 3D.

\subsection{Precomputing Meta-data}
Whenever an image is uploaded, we run pre-processing steps to store meta-data and avoid recalculations. Images are captioned with Florence2 \cite{xiao2024florence}, and images are preprocessed for ControlNets: OpenPose \cite{martinez2019openpose}, DepthAnything V2 \cite{yang2024depth}, Scribble, and Lineart. This metadata is later used for other functions such as search or supporting structure images in some easels.

\subsection{Draw and Paint Workflows}
The Draw easel uses the same workflows for Stable Diffusion XL and FLUX except features for reference images, negative prompt, etc. are not exposed. However, the Draw easel also supports Wan 2.2 as an image generator with our custom workflow.

\subsubsection{Text-to-Image in the Wan 2.2 Model}
The Wan 2.2 text-to-video model \cite{wan2025wanopenadvancedlargescale} uses a \textit{'mixture of experts'} to create visuals. When creating videos, The \textit{'high noise'} checkpoint is responsible for the movement, and the '\textit{low noise'} model creates intelligible high-quality visuals from it. To create an image, we use the \textit{'low noise'} model for a single frame. The Wan model for image generation can render realistic poses and expressions, which is harder to achieve in traditional image models. 

We use the Wan Lightning LoRA which is originally designed for reducing the number of inference steps from 20 to 4 at a classifier-free-guidance value range of 1.0 to 1.5. We found that using the Lightning LoRA at 20 steps leads to much higher quality renders. We added an additional layer of Normalized Attention Guidance at a strength of 11, to enable negative prompts and Lying Sigmas for details. We found that the Lying Sigma value behaves differently in the video model compared to typical image models as it seems to also affect the stylization.

\subsubsection{FLUX Generations with References, Style, and Structure}
Paint works by loading the FLUX model and encoders, then embedding the active LoRAs. We tie three reference images to FLUX Redux\footnote{https://huggingface.co/black-forest-labs/FLUX.1-Redux-dev} using the AdvancedRefluxControl node\footnote{https://github.com/kaibioinfo/ComfyUI\_AdvancedRefluxControl} which adds further controllability. The AdvancedRefluxControl does not have an inactive state, so to avoid switches or automatic rewiring, we modified the code for the node and added a condition to return 0. To control the structure we expose one of the control images from the preprocessed metadata at fixed end percentages as ControlNet using ControlNetUnion \cite{zhao2024uni} (pose: 0.9, depth 0.7, and lineart 0.4). We also use Normalized Attention Guidance for negative prompt and Lying Sigmas for details. The generation is done with 8 inference steps leveraging FLUX Turbo\cite{FluxTurbo}.

Because FLUX Redux impacts the conditioning, image references affect both style and composition which can conflict with ControlNet for structure. We created an alternative version of the workflow where one of the image references ties to ByteDance USO \cite{wu2025usounifiedstylesubjectdriven} to define only style while preserving the compatibility with Redux and with ControlNets.

\begin{table*}
    \centering
    \includegraphics[width = 0.85\textwidth]{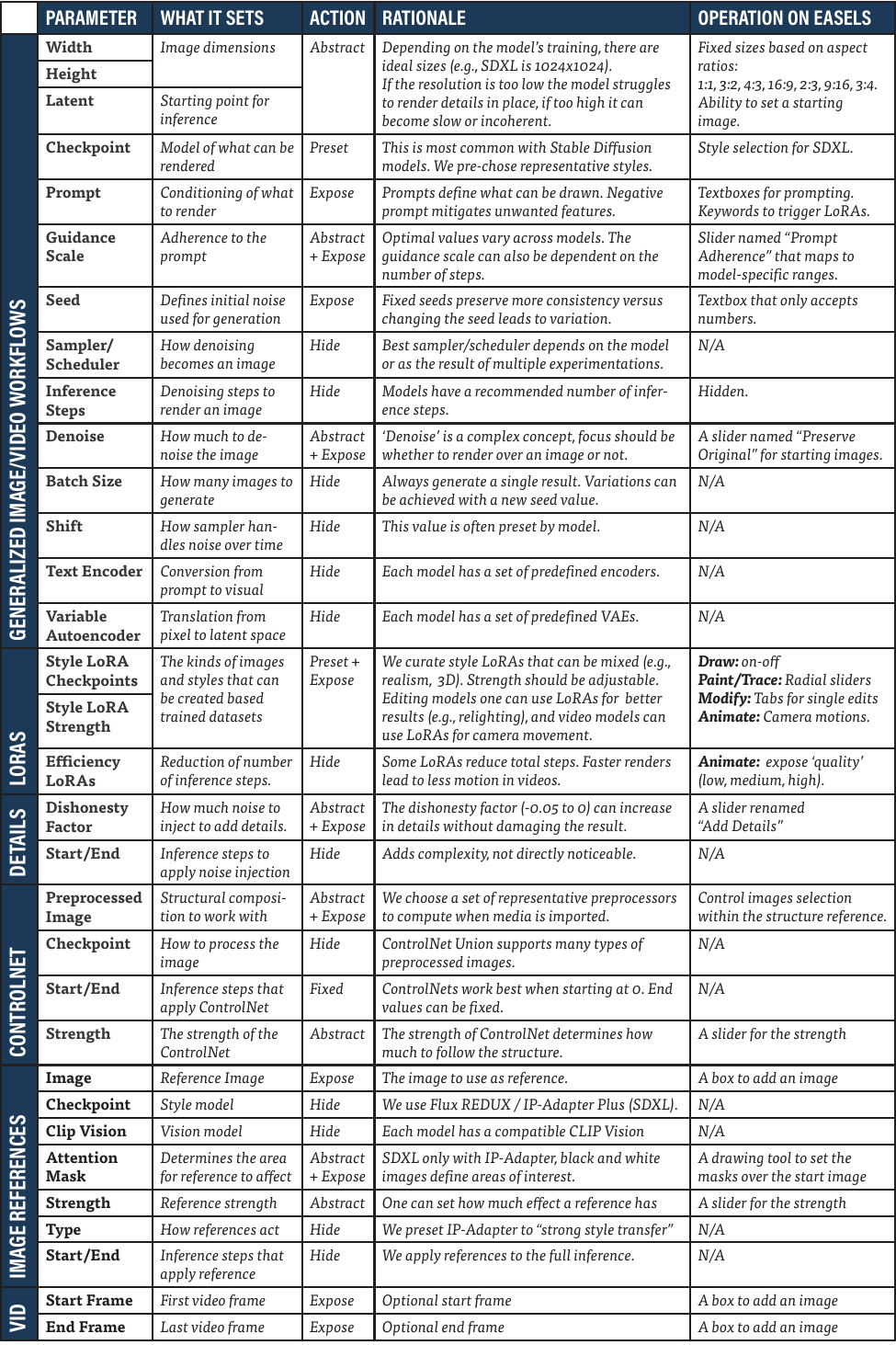}
    \caption{Rationale of workflow parameters as they get converted into easels.}
    \label{fig:rationale-table}
    \Description{}
\end{table*}

\subsection{Modify Workflow}
Modify uses Flux Kontext to edit existing images, such as changing cinematographic attributes. We add a set of LoRAs for changing camera angle, relighting, and styles. Unique to this workflow is the ability to recompose an image into a completely new aspect ratio.

\subsection{Trace Workflows}
We use two workflows to support Trace - one for FLUX and one for Wan 2.2, both based on workflows and custom nodes by Github user Log(td)\footnote{https://github.com/logtd} with a technique called FlowEdit \cite{kulikov2024flowedit}. FlowEdit reverse-engineers a tagged starting image and reconstructs it based on a new prompt. This approach came about before editing models such as FLUX Kontext and Qwen Edit, and thus can transform images in interesting ways. The FLUX workflow is based on Log(td)'s Fluxtapoz node package\footnote{https://github.com/logtd/ComfyUI-Fluxtapoz}. We modified the sample workflow by adding our set of custom LoRAs, ControlNet, and USO style images \cite{wu2025usounifiedstylesubjectdriven}. In our implementation we use 20 inference steps.

For FlowEdit in Wan 2.2 image generation, we created a workflow combining our text-to-image workflow with a workflow inspired by Zack Abrams\footnote{https://tinyurl.com/AbramsWanFlowEditWorkflow}, who modified a Hunyuan Video FlowEdit workflow\footnote{https://github.com/logtd/ComfyUI-HunyuanLoom} for the Wan 2.1 image-to-video model. 

\section{Atelier in Extended First-Use}
\revision{We conducted a first-use exploration with creative professionals using the system for 4 hours (including breaks), targeting an open-ended session that could accommodate the learning curve. Our goal was not to evaluate the performance of Atelier, rather to see how creative professionals might appropriate a tool like Atelier. This merited a study design mixing walkthrough demonstration and observation \cite{ledo2018evaluation}. Specifically, we wanted to learn how practitioners might use Atelier, different strategies they might adopt, how it might fit existing creative processes, expressiveness, including threshold and ceiling \cite{myers2000past}), etc.}

\subsection{Participants}
\revision{We recruited 5 creative professionals (2 female, 2 male, 1 gender non-conforming) aged 29-31. We followed a purposeful sampling recruitment \cite{blandford2016qualitativehci} and reached out by email to a mixture of direct contacts. Their discipline of expertise ranged in a variety of areas within the media and entertainment industry. Each participant had using generative AI in their creative practice. As a token of appreciation for participating in the study, participants received the equivalent of a \$400 USD gift card. The procedure was approved by institutional ethics review boards prior conducting the study. Because our aim was not generalizability but rather to surface diverse strategies of usage, 4 hours of in-depth rich interaction per participant produced substantial information-rich data. Participant background and AI experience is summarized in Table \ref{tab:participants}}

\begin{table}
\caption{Summary of Study Participants showing their background and experience with AI.}
\label{tab:participants}
\includegraphics[width = \columnwidth]{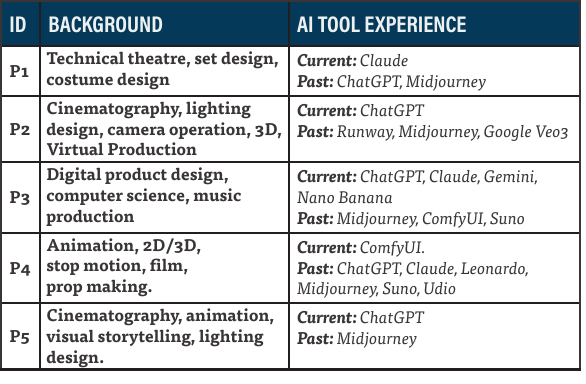}
\end{table}

\begin{table}
\caption{Survey responses to paired questionnaire for current AI tools and Atelier. The third column shows the delta between responses for each individual participant.}
\label{tab:user-study-survey}
\includegraphics[width = \columnwidth]{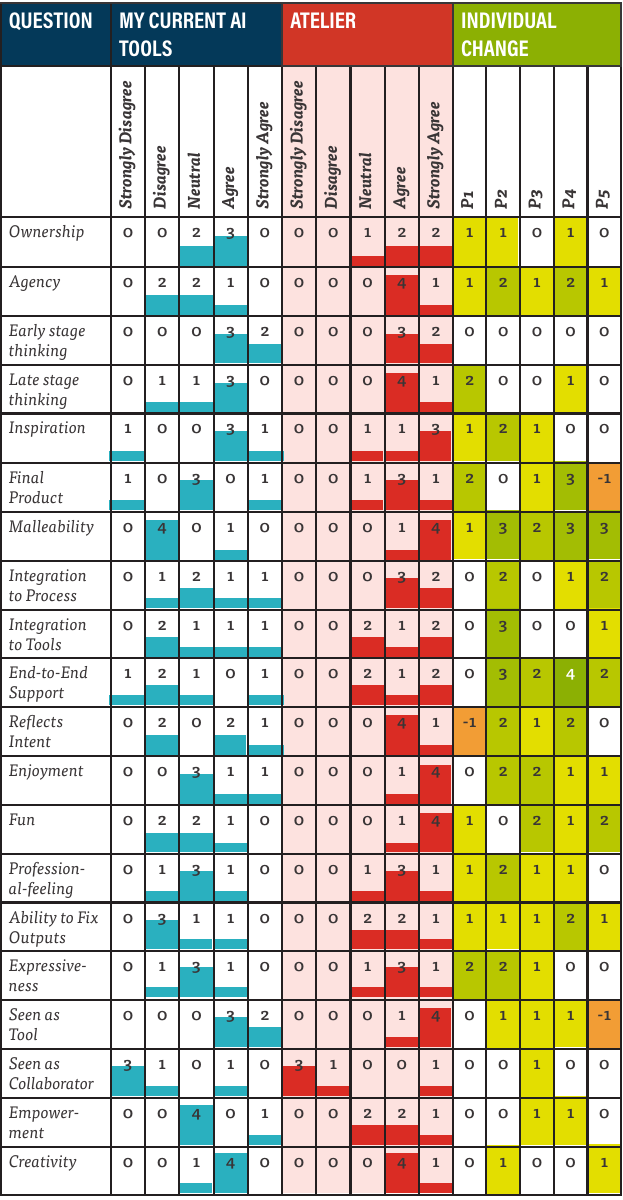}
\end{table}

\subsection{Procedure}
\revision{Participants received an email invitation for the in-person study, and provided with the video figure of Atelier. Participants were asked to think of a project in advance but were not expected to conduct extensive preparation. The study consisted of three core activities: pre-study interview, guided demo and free-form exploration, and post-study survey. }

\revision{\paragraph{\textbf{Pre-Study: Semi-Structured interview (15-20 mins).}} After consenting to the study and screen-recording, we conducted \textit{pre-study semi-structured interviews} to learn participants' background, how they use inspiration material, their AI experience and feelings of ownership, etc. This pre-study interview took 15-20 minutes.}

\revision{\paragraph{\textbf{Guided-Demo with Free-Form Exploration (\~3.5 hours).}}
The experimenter provided a computer, display and peripherals. Participants used Atelier through a web browser with full access to the experimenter's device, including web browser and creative applications offline to access familiar tools if/when needed. }

\revision{The experimenter was available for lightweight guidance to reduce friction. Participants led the exploration - features would not be introduced until they had either stopped testing the feature or using past features in combination. The experimenter would ask questions or gather impressions through think-aloud (unprompted).}

\revision{Participants were introduced to the canvas navigation. They were then shown quick operations. Participants would import materials as they were introduced to different easels, starting with 'Draw' given its simplicity. The subsequent order would depend on the individual's project. Lastly, participants were exposed to the provenance, organization, and sense-making tools. Participants continued using the system freely for the remaining time. Participants were encouraged to take breaks, most taking 2-3 breaks throughout, including one longer 15-20 min break half-way. }

\revision{\paragraph{\textbf{Post-Study Survey (10 mins).}} The last 10 minutes were allocated to a\textit{post-study survey} where participants entered their demographic information and answered questions about their practice, as well as a set of paired Likert-scale questions contrasting their current use of AI tools with Atelier. An example question is: \textit{"I feel a sense of ownership with what I create with [Atelier | my current AI tools]"}, see Appendix \ref{sec:questionnaire} for full details.}

\subsection{Data Analysis}
\revision{All sessions were screen and audio recorded with participant consent, which were transcribed with filler words removed. The qualitative data was analyzed jointly between the first and second author. The first author conducting open-coding on the transcripts. The second author, who conducted the experiments, reviewed the codes and data. Through iterative discussion and thematic clustering \cite{charmaz2006constructing}, the authors identified a set of themes. The post-study survey Likert-scale data was used to complement the findings and analyze the changes in sentiment across individual participants.}

\section{Results}
\begin{figure*}
    \centering
    \includegraphics[height=\textheight]{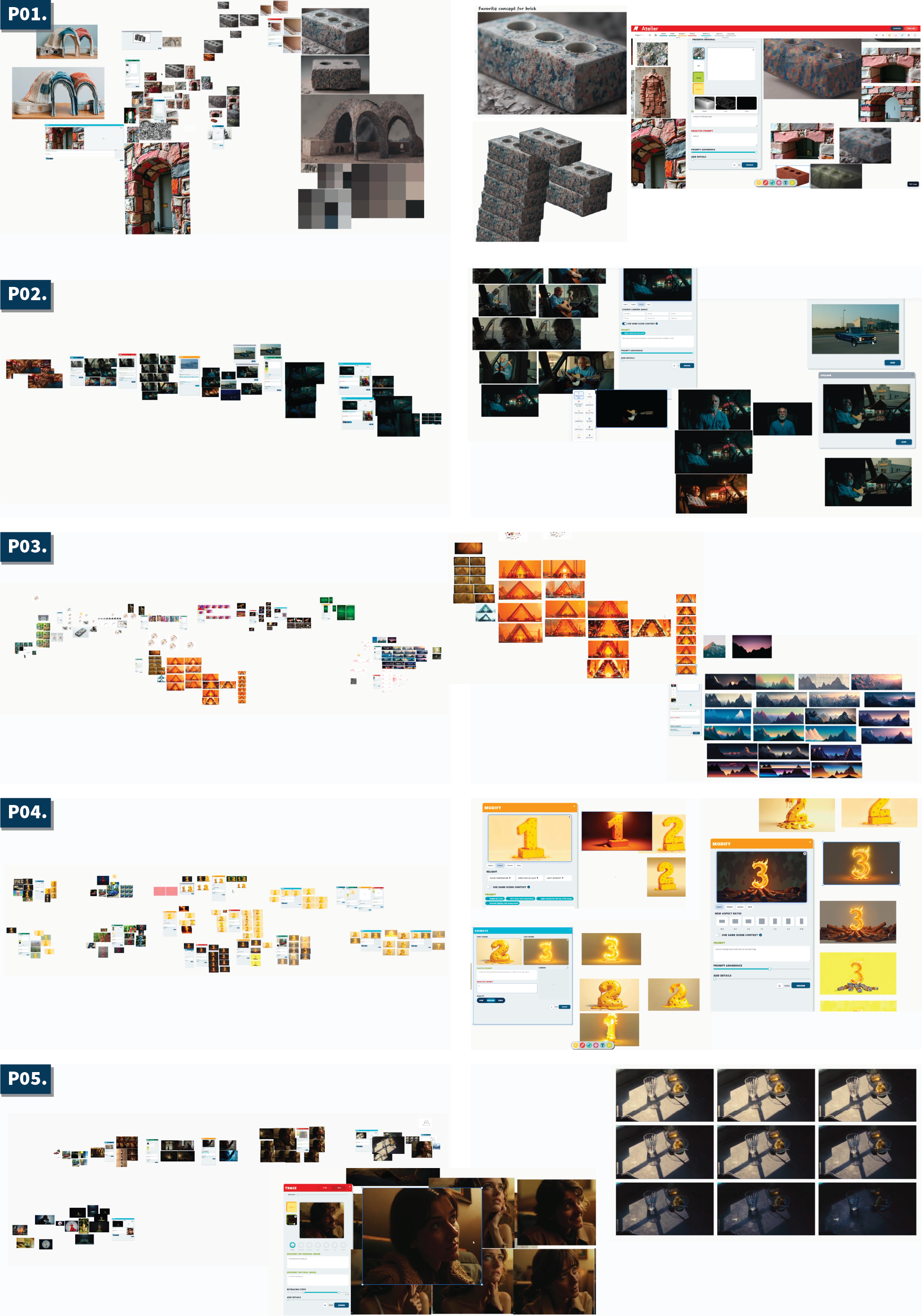}
    \caption{Canvases created by participants using Atelier. P1 shows a concept for a brick made out of fabric scraps; P2 shows shots for a scene of an old man playing a guitar inside a truck; P03 created a variety of concepts, including an app for visualizing stock trading as landscapes; P4 recreated an old animation exercise of numbers made of different materials morphing; P5 explored different shot concepts for a film and then for a commercial.}
    \label{fig:results}
    \Description{}
\end{figure*}

\revision{Participants all created canvases using the different functions and easels (Figure \ref{fig:results}). P01 worked to materialize a "brick made of clothing scraps" for set design work. P02 created a set of film scenes of a man playing a guitar outside of a casino, iterating on lighting, camera angles, and camera movement. P03, after open ended exploration returned to an old app idea that rendered stock graphs as mountain ranges and used Atelier to prototype the visual translation. P04 recreated an old animation project of numbers morphing into different shapes as a movie countdown. P05 explored shots for a film concept and then built out product shots for an advertisement project, iterating on light, framing, and movement.}

\subsection{How was Atelier used?}

\revision{\paragraph{\textbf{Favourite Easels.}} Participants all described their favorite easels. P1 and P4 appreciated Paint the most. P1 liked the negative prompt and structural reference as it enabled them to have more control of the composition, whereas P4 noted that it gave the most flexibility and options to explore. P3 and P5 liked Trace the best as they felt they could transform starting images: \textit{"I think it's because I have this retracing steps as like a control. Mechanism for adjusting the severity in a bit more of a predictable way. I feel like the retracing steps I could become like a power user of [this easel]... This gave me more predictable results"} (P3). P2 saw the strongest affinity for Animate: \textit{"the ability to see it in video, see it in motion gives me so much to think and to reflect creatively... so much as a possibility [compared to] an image [as] extra step of my creative process... it changes everything"}.}

\revision{\paragraph{\textbf{Biggest Challenges.}} P1 and P2 described Trace as the most challenging easel to work with. P2 was unclear as to how it worked despite having achieved some successes: \textit{"what are we actually trying to do here? Because we have the original image, we have final image..."}. Having to describe the original image and then having to describe the final image was something that all participants were initially confused about. For all participants, prompting was a major challenge. P1 and P5 resorted to using Claude - P1 would ask for better phrasing of the prompt, whereas P5 would use it to find the right words to describe something or to translate to English from their native language. P2 noted how often with generative AI the only way to obtain variations is to \textit{"write better, just give a better prompt... it's very limited of what you can do"}.}

\revision{\paragraph{\textbf{Clarifying Design Intent.}} Atelier provided participants with more means to achieve the visual they wanted. Part of it lies on the experimentation, with P2 "lik[ing] how precise [one] can be with the program... options of Draw, Paint, Modify gives you more room to experiment". P1 described how: \textit{"it gives you control... it's easier to know how to imagine... this is closer to what I want"}. Creating different images often served to have a better sense of what the design intent was: \textit{"it's nice to see other explorations... Now I am liking more this idea than [a previous idea]" (P5)}. Undesired results also helped fine-tune the design intent, as participants would have a better sense of what they do not want, thereby reducing the ample space of possibilities:\textit{ "for me it would give me clarity... I don't want the patchwork brick, I want the scrapped brick. So then I'm gonna shift how I'm gonna build things so it can help me think, having clarity of what I don't want. It's faster not to get lost"} (P1). }

\revision{\paragraph{\textbf{Support for Different Stages of the Creative Process.}} The post-study survey suggests that participants can envision using Atelier for inspiration, early-stage thinking, late-stage thinking, and final products more than their current AI tools (Table \ref{tab:user-study-survey}). However, perspectives were more nuanced, depending on the practitioner's primary medium and their goals, and was primarily seen as a conceptual tool. P1, despite having used AI images for a final project in the past, feels that generative AI is conceptual because it remains digital. She sees her set designing as a physical hands on activity: \textit{"I would use images to visually communicate. Take the concepts take the ideas, take my interest, and communicate and then arrive at a thing that people are like 'yes, that's what we're building'"}. P2 noted how Atelier would enable them to create more realistic pitches: \textit{"I see [myself] using [it] for pre-visualization from every stage... thinking about things and brainstorming... reference videos or have reference images for a pitch package or a grant application. I see myself using it for pitching to a director with this like saying, oh, this is my vision, this is how I thought about your scene. This is the type of movement and tone I want being very specific about it and probably landing that director, getting that director to hire me because it's going to be so specific that it's kind of hard to beat when you show something that is unique versus an image of a movie that already exists"}. P2 also said they might use Atelier to create full videos for fun. P3 saw value in concept generation, but also saw their easels as pipelines they wish they could get API access for: \textit{"what I would love to do is create this pipeline and then ultimately export it or serve it as an API for my own apps"}. P4 and P5 saw value in using Atelier for storyboarding or animatics, P5 especially noted it being useful for \textit{"commercial stuff... they really want to see how it's gonna look. And if you're bringing something like this, you're going to save a lot of problems with them during the shoot"}. }

\subsection{How did Participants Organize the Space?}
\revision{\paragraph{\textbf{Canvas Layout.}} Participants developed their own ways of moving through the canvas as they worked. Every participant had an element of working left to right to varying degrees. P2 and P3 organized their canvas straightforwardly chronologically from left to right, with P2 explaining: \textit{"I tried to think chronologically... moving the way that we read in the Western world from left to right, and I was just trying to keep that consistent"}. P2 noted that if they were working on a longer term project they would create separate corners for each easel. P3 moved left to right temporally and arranged variations vertically. P4 kept outputs on the right and references on the left closely to the easels, noting: \textit{"If I was organizing it with more time, I would be a bit more separated into more sparse clusters so that I have more space around it to work"}. P1 moved from the center outwards, placing preferred work to the far right and making those images bigger for emphasis, with discarded material on the far left. P5 moved from left to right, then started a new row below the current work when exploring a new topic. Most participants (P2, P3, P4, P5) clustered outputs closely around the easels that generated them, with P4 noting that even when zoomed out: \textit{"I can visually tell what's what... little cluster over here, little cluster over here... I can tell that that's different things going on."} Participants took different approaches to managing the volume of generations. Most (P1-P4) kept the majority of their generations, deleting only outputs that were completely off from their targets, whereas P5 deleted both easels and generations more liberally. Several participants (P2, P3, P4) used the grouping and packing tools to spatially lay out their work, while P1 and P5 were more comfortable with overlapping elements on top of each other.}

\revision{\paragraph{\textbf{Easels as Spatial Anchors.}} Easels functioned as anchor points within the infinite canvas. P1 articulated that with \textit{"canvas based work, it's so huge, you can get so lost. I think the easels provide you some structure without limiting you... what the easel does is... it helps you to organize things... so you can work within the flow happening, but then you can also organize so that when you're back, you don't get overwhelmed by the amount of stuff"}. They explained this would allow them to come back to work on a new day with a \textit{"fresh place to work while still being able to draw connections and go back on the work"}. Participants valued the space of the infinite canvas with P2 describing: \textit{"any canvas style creation, it's almost like an infinite. It's a multiverse. You can create a multiverse of ideas and options... I like how it feels endless"}. They noted that even with more erratic organization, they \textit{"could still see the process"} spatially. Participants sometimes treated easels as personalized workspaces they configured for their own workflows. P3 described easels as: \textit{"I see this as my little factory area... my playground... this is my space... there's so many different parameters and ways to set things up. I don't want to even slightly forget how I did it. I like to keep that persistence. It's almost like I built the tool because I added all these parameters in it"}. P2 envisioned organizing the space by easels with extended use: \textit{"if I was gonna work this more seriously, I would probably separate a bit more tools. This is my Drawing corner. This is my Modify corner"}.}

\revision{\paragraph{\textbf{Provenance Tools.}} The other provenance tools such as timeline, history, heat map, lineage panel and recreating easels options were used to navigate and make sense of their work and the process it took to get them there. P2 called the history feature \textit{"life changing... why doesn't everything have this, like a history, a visual history of what you're doing"}. P3 explained \textit{"I like to preserve the history of what I've done. It both makes me feel like I'm making progress and I can see where I've come from, revisiting old ideas when maybe I've forgotten about them"}. The "recreate easel" feature was used to return to specific parameter configurations with P1 describing it as \textit{“the gift of being given your image and then the notebook that shows you how you got to the image... you can go back to your sketch of the painting easily"}. P4 valued how the provenance system addressed reference management challenges noting it is \textit{“really useful. It's good because then it safeguards your process. It’s dummy proof... the management references... you [might] have to go to an outside software... but then you have it within the software here, which is handy. And then you make sure that you don't lose it because it's in your project"}. The preservation of their process was important to participants both for managing references as well as for making creative connections and reflections. P3 described \textit{“I like to preserve the history of what I've done. It both makes me feel like I'm making progress and I can see where I've come from revisiting old ideas when maybe I've forgotten about them. I think that’s pretty important”}.}

\subsection{How Does Atelier Compare to Other AI Tools?}
\revision{When using Atelier, participants brought up comparisons to previous experiences with AI image or video generation platforms. This included prompt-based platforms such as ChatGPT and Midjourney, as well as node-based platforms like ComfyUI. }

\revision{\paragraph{\textbf{Comparison to Prompt-Based Interfaces.}} Compared to prompt-based interfaces, Atelier’s multi-modality offers visibility as to what the models can do through the easel's design (e.g., via the slots that can be filled for image references) (Table \ref{tab:user-study-survey}). To participants, language-based approaches do not provide clear means to create with generative AI: \textit{“[one has] to communicate to a computer only through words, which is limited to represent what [one has] in [their] brain. Instead of allowing the artist a little bit of autonomy with the tool, it makes [one] go through that [language] layer... how can you represent what you have in your mind physically if you don't have all the tools available to you, if you don't have a hammer, if you don't have a chisel...”} (P4). }

\revision{Compared to prompt-driven interactions, Atelier shifts to working with the material that changed participants' relationship to AI generation from talking to a model to manipulating content: \textit{"When I go in Midjourney or Runway, I feel like I'm interacting with the AI... I'm asking something from it and receiving that image. Here, I feel like I have creative control. It's my creative workspace and not the AI's workspace"} (P2). P5 noted that in comparison to Midjourney, Atelier provided starting points: \textit{all these options like Draw, Paint, Modify... feel more friendly. You have more options to start... that's the most intimidating part for me when I am trying to create something"}. This shift gave participants agency when generations failed, rather than blindly rewriting prompts, they had \textit{"accessible knobs"} to adjust. }

\revision{P2 felt that when using Atelier brought focus to the process itself: \textit{"it's a creative process even to interact with the system. While when I'm working with other video generative tools, it doesn't feel like it's creative"}. P2 also suggested that many details, from the vibrant colors making the system feel approachable, to the choice of naming conventions in Atelier encouraged creative thinking: \textit{"what I feel is creative is the tools here. Even the way you're naming the tools, like drawing, painting, sketching, collage... it's mentioning art forms, and it's encouraging me to think in art in a more creative way rather than prompting}".}

\revision{\paragraph{\textbf{Comparison to Node-Based Systems.}} P3 and P4 both had prior experience with ComfyUI. They articulated that Atelier was more accessible and intuitive to learn and use. While node-based systems offered more fine-grained control, the learning curve served as an intimidating barrier to creation, where many participants saw the software was a means to an end and if the \textit{“learning curve to the software is as big as the art I'm gonna make, I don't want to do it”} (P1). P3 recounted that tools like ComfyUI: \textit{"are very daunting tools.. the different inputs and outputs, little connectors and knowing where they go and how they operate... there's just so many... even coming from the software development world and using Blender, I haven't fully adapted to nodes... I tried [ComfyUI]... and I failed because it was too hard of a learning curve"}. In contrast, Atelier: \textit{"at least allows me to experiment with these pipelines in a bit more of a novice way”}. P4 noted that Atelier\textit{"feels a lot more personal... more collaborative... It feels more visual, it feels easier to explain, feels easier for even somebody who perhaps doesn't have experience using tools. If I were to do this in front of them, then visually they can already understand something, versus with ComfyUI you have all this text and nodes... and then finally an image at the end. With [Atelier], it's only images. It's easier to show this [easel] is generating all these images... versus a whole system generating one image”}. This made the work legible for participants' own understanding and further for potential collaborative sharing. P1, often working with less computer-literate collaborators, noted that\textit{ “there is something about this that makes it easy for people to use that you don't need a lot of buy-in on how this could be useful. You get the results as you're working through it"}.}

\revision{The increased speed at which participants could get desired results allowed for context to be maintained during creative exploration. P4 noted that \textit{“it would have taken [them] so much longer to get to a result like this in Comfy”}. With all the generations in one place, Atelier provided a continuous visual history, which P4 referred to as a \textit{“fresh memory”} of how ideas developed. Moreover, because the iterations were close together in time and space, they could \textit{“trace it back mentally”} and \textit{“see it all there”} by zooming out. P4 noted that if they used ComfyUI, their generations would have required them to split the work across two days. This continuity enabled forms of comparison and judgment that were difficult in chat-based prompting, which P3 described as \textit{“rigid”} and \textit{“utilitarian”}, oriented toward reaching a single output rather than exploring alternatives: \textit{“this tool allows me to basically explore a search space of creative possibilities... whereas other AI tools I've used don't really use this canvas approach... feels more task focused”}.}

{\subsection{How does Atelier Support Craft and Ownership?}
\revision{The pre-study interview highlighted some base beliefs from participants regarding their views on craft and ownership. Yet the use of Atelier also highlighted new ways to reflect people's craft and ownership. Specifically, we identified themes of intent, malleability, collective outputs, and the participants' ability to understand AI as a medium. The post-study survey results (Table \ref{tab:user-study-survey}) also highlight elements relevant to these changes, namely the categories of: ownership, agency, malleability, reflecting intent, ability to fix outputs, expressiveness, empowerment and creativity. All of these categories saw mostly sentiments in favour of Atelier.}

\revision{\paragraph{\textbf{Intent.}} One way in which participants felt ownership is through intent. In the pre-study, P02 described how their generations are often unique because they \textit{"have very strict ideas [their] mind and they take so many different prompts"}, while P03 described how in the past they made an AI comic where they felt they were \textit{"creating that art form"} because they were \textit{"being very intentional"}. The usage of Atelier also demonstrated how intent was reflected. P01 described: \textit{"It feels like something I had in my head before and AI just made... the image generation is just a tool... to making sure that people understand and visualize the same thing as I visualize"}. P05, while unsure about their sense of ownership, felt Atelier enabled experimenting to refine their design intent: \textit{"this thing already has some ideas that I have or a mix of ideas that I want to reach}".}

\revision{\paragraph{\textbf{Malleability.}} Malleability showed the most considerable increase across participants in the post-study survey. In the pre-study interview, P01 described how they need to \textit{"have a lot of input into [AI] to then feel like it can be [theirs]"}. P02 described how prompts are achieved from many iterations: \textit{"I crafted them like a diamond... the diamond was always there but then it needs to have that cute shape. Otherwise it's just a rock."} These points highlight how each generation leads to adjustments in parameters for rapid iteration. After using Atelier, P02 found more nuance in how they see ownership \textit{"here I feel like I'm a creator. I'm creating, combining, sketching, drawing... it's about the process... prompting feels a little bit more technical... You're trying to guess formulas and what the AI thinks and try to break those formulas"}. P03 noted how easels embodied a set of parameters that enable manifesting their design intent: \textit{"[in] getting [the concept] accurate and right... the pipeline to achieve this technological possibility ability is that there was craft in that."} This sentiment was echoed by P04, given how in tools like ChatGPT \textit{"you have to write a whole paragraph of what you really want. But then that might not be the result".} In contrast, \textit{"with [Atelier] you isolate all those characteristics and then you can make sure that all those aspects stay the same even after you modify other things."}. }

\revision{\paragraph{\textbf{The Collective Outputs.}} The malleability aspect considers how the generations lead to new settings and iterations. This process can also be looked at from the point of view of the outputs generated. The set of outputs, over time, can reflect the changes in settings as generations went in specific directions. P02 already had described the iterative process to get the right prompt. However, since Atelier places all of the outputs within the canvas, P03 noted a sense of craft and ownership from seeing the different generations side by side, reflecting on the spatial arrangement and the emergent results - their sense of ownership and craft comes from seeing the journey: \textit{"even though the source material is from a very famous film, as we were going along... picking up different variations and worlds... I could start to pick and almost make opinions about what direction I wanted the style to go... And I think part of it is a having the control to guide things much easier.. But also seeing the variations, seeing them all laid out in this way, being able to accept and reject and ultimately make it pick a creative direction from a large set of explorations... is part of that creative process that gives me ownership"}. P04 tied the feeling of using Atelier to storyboarding \textit{"it feels more like storyboards and like thumbnailing"}.}

\revision{\paragraph{\textbf{AI as a Medium.}} When discussing why they do not feel ownership over their AI generations, P05 explained it was because they do not understand how generative AI works, in particular they stated: \textit{"It's a challenge sometimes... I get desperate really easily. Like, that was not my idea and I just don't want to take [the] effort"}. P05's use of Atelier appeared to create a sense of control and agency, especially in their use of the Trace easel, enabling them to generate similar shots to their reference material but better aligning to what they are looking to create. While the feeling of ownership did not change, P05 developed more confidence in their experimentation. This is also reflected in their post-study survey response, with increases in agency and ability to fix outputs. P04 described similar importance to understanding AI as a medium: \textit{"It's like accepting what it is and using what it can give us to shape what your imagination is trying to do in a way... it's similar in to what artists do physically too. I mean, at the end of the day, what you imagine is not necessarily what you end up making. There's a lot of things in the path of getting something done from beginning to end that can change"}. While we did not explicitly ask about understanding as a medium, the sentiment responses to the post-study survey (Table \ref{tab:user-study-survey}) show a change in sentiment in a few metrics that could be associated to understanding AI as a medium: malleability, agency, ability to fix undesired outputs and sense of empowerment.}

\begin{figure*}
    \centering
    \includegraphics[width = 1.0\textwidth]{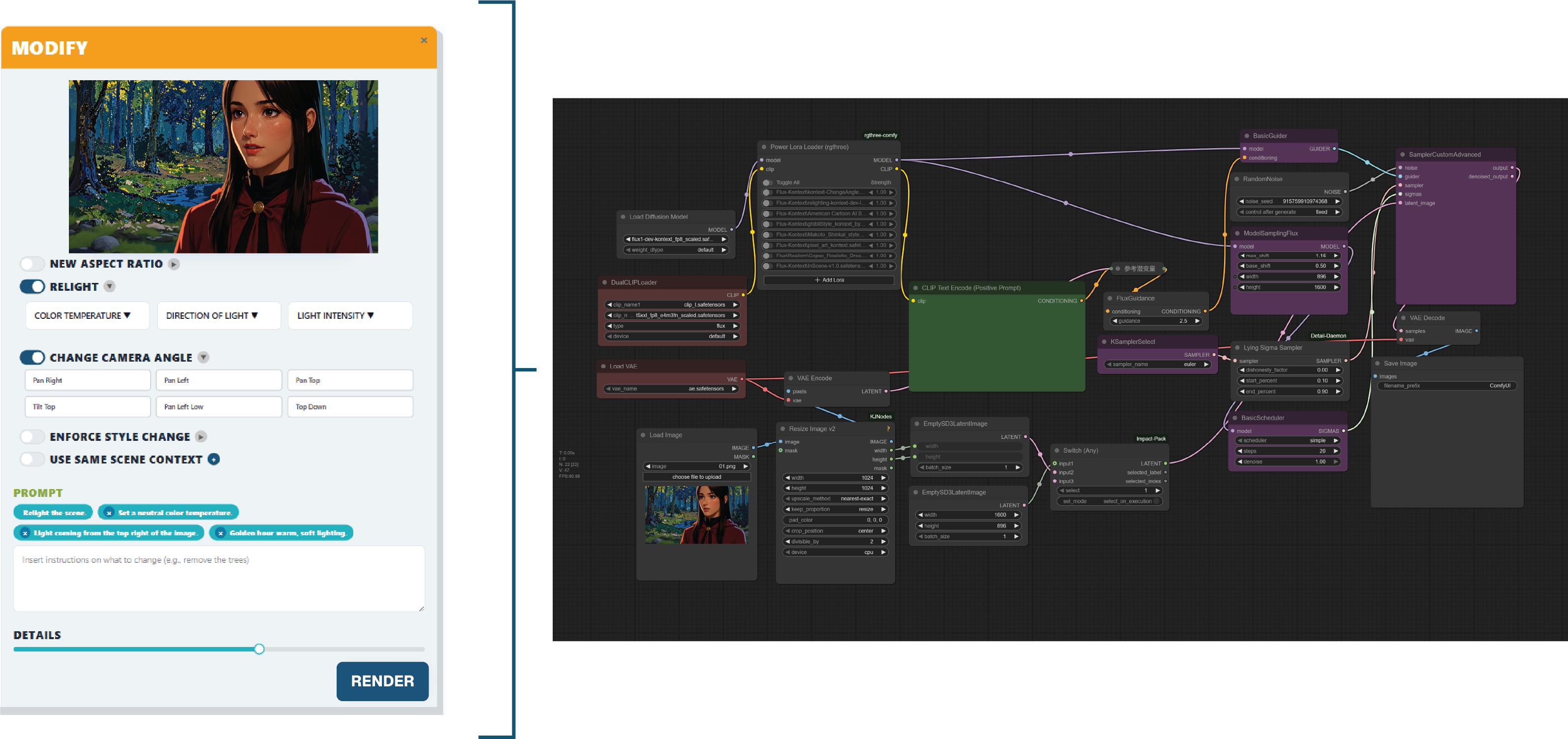}
     \caption{Example of the Modify easel and how it maps to its corresponding ComfyUI Workflow.}
    \label{fig:easel-to-workflow}
    \Description{}
\end{figure*}

\section{Discussion}
Protosampling as a concept provides a lens with which to look at the creative process to inform the design of authoring systems. Atelier represents one way of doing this, acting as a way to validate the concept \cite{wiberg2014makes}. As part of the research process, using Atelier instigated critical reflections, from it being a means to Protosampling, to the challenges and opportunities of generative AI being non-deterministic, to the conceptual implications, technical and conceptual limitations, and potentials for future research.

\subsection{Atelier as a Means for Protosampling}
Early in this paper, we proposed Protosampling as a method that balances the curatorial nature of sampling along with its extraction of insights, to the act of prototyping which manifests ideas. This promoted many of the design decisions made in Atelier.

\subsubsection{Blending Thinking and Making }

The media-first approach mirrors the real world \cite{dourish2001action}, where references and inspiration are dealt with as wholes as the creator may not know why something is useful yet \cite{kochImageSense, luceroFramingAligningParadoxing2012, lucero2012moodboards}. Easels, together with the quick operations, act as the key way to enable the 'making' part of the rationale. \revision{Many of our participants described that the interface was very intuitive, where the space felt like it could be called theirs.} We also found that Atelier directly supports the procedural actions in Protosampling, particularly derivation and combination approaches. Aspects such as analogy, metaphor, and first-principles are processes that are much more complex and more up to the practitioner to devise as part of their creative problem solving. This might be something where creators might find more value directly engaging with other people or learning more about the world around them. \revision{Importantly, every participant emphasized that working in Atelier felt playful or fun. This sense of playfulness appeared alongside behaviours such as freely branching, recombining, and testing ideas, suggesting that the Atelier supported exploration that felt open and low-stakes. }

\subsubsection{Encapsulation}
With current approaches, creators have to decide between two options. One is to use a dedicated service, such as Midjourney \cite{midjourney} or Runway \cite{RunwayML}, and use the interface and options provided. These can be fairly accessible, though the scope of each service varies, which often means work has to migrate across services. The other approach is to use a system where they can access the models directly such as ComfyUI or a gradio-based interface such as Automatic1111\footnote{https://github.com/AUTOMATIC1111/stable-diffusion-webui}. However, the challenge, as shown with participants who felt frustration with ComfyUI or node-based canvases which take time and training to learn, becomes how to harness the potential of these node-based interfaces and make them more accessible. This is where the potential for designing interactions comes in. This motivated mapping Draw and Paint to enhanced text-to-image and image-to-image workflows, Trace to the use of FlowEdit, and Modify to editing models such as FLUX Kontext. The selection of model, checkpoints, LoRAs, etc. comes second and is prepackaged. We also thought about the preprocessed images. In early ideation, we intended having the depth, pose and lineart images for ControlNets as media within the canvas. Yet, we realized at some point that these were only being used in the easels that support structure references. We realized that having these images would flood the canvas with preprocessed materials, and would require users to think about what they mean and how to use them. Having them as suggestions in the structure compartment of the easel removes that complexity, though it still requires acquiring the literacy to know what these images are and what they do.

One example of how an easel maps to a workflow is the \textbf{\textit{Modify}} easel (Figure \ref{fig:easel-to-workflow}). Besides using FLUX Kontext-dev and LoRAs for relighting, camera changes, and styles, we extracted potential keywords from the different LoRAs and added them as prompt pills. The toggles on the interface map to different LoRAs or to the switch for aspect ratio. 

These are some examples of the many considerations when building these abstractions, looking to have low thresholds and high ceilings \cite{myers2000past}, provide a set of building blocks \cite{ledo2018evaluation}, and expose the right parameters so that expressiveness can match the task as much as possible \cite{olsen2007evaluating, palani2022don}. In building these abstractions, we considered experiences from different online communities from daily engagements such as Reddit\footnote{https://www.reddit.com/r/StableDiffusion/} and Banodoco\footnote{https://banodoco.ai/}, as well as existing documented first-person accounts \cite{ledo2025generative} and our own experiences learning and using these technologies. \revision{One participant (P3) stated that after using Atelier for the period of the study, they felt that they understood ComfyUI more. This points at a possibility of progressively introducing more advanced concepts or unlocking more complex features in the system, and could further explore creative systems as learning grounds for AI concepts through use.}

\subsubsection{Organization and Provenance}
From participant use, content often congregated around easels. In that sense, the easels can act as anchors for content. At the same time, the canvas metaphor makes it so content is scattered without structures, even without piling as an option. Given the discussions on knowledge work, storage and search, and material retrieval \cite{bondarenko2005documents, bondarenko2010requirements, sellen2003myth, henderson2009empirical}, we saw high importance in being able to explore the content and create more permanent collections. We believe that engaging with these externalizations can invite reflection \cite{dix2011externalisation, schon2017reflective}, as now the entire set of materials exist as triggers for reinterpretation \cite{eckert2000sources}. \revision{This suggests that media-first and freeform creation can be viable, but an infinite canvas needs to be paired with equally powerful sensemaking features for participants not to get lost. Creative work is often very collaborative, and participants had others in mind that they wanted to show the work, whether it was clients or collaborators. The space lends itself well to collaboration or sharing of process, as participants felt that it was easy to understand their own journey.}

\subsection{The Two Sides of the Coin of Open-Endedness and Activities}
One challenge with Generative AI is the lack of determinism. On the one hand, this can be a powerful advantage, as there are many ways to solve a problem, providing many paths of least resistance. For example, consistent characters remain a challenge with AI image generation, often requiring training custom LoRAs. With the \textbf{Modify} easel, one can get some alternative camera angles for the character which can be brought into a starting frame for a video. Another option is to use \textbf{Animate} to generate a video with a start or end frame to prompt for character or camera motion, then extracting poses from those frames. These multiple pathways to solve a problem, together with the ability to remove backgrounds and reassemble in a collage means that it is possible to achieve multiple character poses and positions and find new ways to reuse them. \revision{In the user study, we also found novel uses of the Collage easel. P1 created multiple copies of a brick to think about how to create a doorway, while P2 used Collage to create composite shots that could be brought into Trace. Collage was especially useful when other easels were not following their exact intent, such as rendering the wrong truck.}

The open-endedness can also be a problem, especially when the extent of a model is not fully understood. Terminology such as 'denoise' or 'schedulers' can be difficult for novices to pick up, and these are terms that exist outside of most creative fields. While there are going to be new terms that need to be picked up, how does one learn the best time to use FlowEdit versus Flux Kontext? For example, both can restyle an image, but FlowEdit can have a lot more creative freedom. The reality is that even with experience, it is not a straightforward answer, and it is one that invites failing multiple times and learning from those failures. However, this also leaves opportunities for surprise even when understanding how to use these techniques.

\subsection{Conceptual Implications}
While the primary concern within Protosampling is the focus on procedural action and the trail, the reflection resulting from this activity has a direct impact on the process itself. 

\paragraph{\textbf{Transformation.}} The first conceptual implication of Protosampling refers to the transformational nature of a reflective practice \cite{sawyer2024explaining}. Evaluating what has been collected or created by a certain point will inevitably reframe the creative problem. Ohlsson \cite{ohlsson1992information} claims this can take place by elaboration (representation changes from adding information), re-encoding (rejecting part of the original interpretation of the problem and revising it), or constraint relaxation (changing an incorrect representation of the goal).


\paragraph{\textbf{Emergence.}} The second conceptual implication is the emergent nature of the creative process, which Cross describes as unrecognized properties that are integrated into future concepts \cite{cross1997descriptive}. This highlights the interplay between being intention-bound (aligning to original plans) and emergence-driven (openness to external influences, learnings and inspiration) \cite{gaver2022emergence}. Ideas do not exist in isolation \cite{cross2011understanding}, instead they are all part of a connected, directed and rational process \cite{weisberg1993creativity, weisberg1986creativity}, one in which externalizations continue to invite new follow-on ideas \cite{sawyer2021iterativecreativeprocess}. 
\revision{
\paragraph{\textbf{Bias and Fixation.}} An interesting question is whether the use of generative AI can increase bias or design fixation. All of the models have different biases and optimizations. With Atelier we support a variety of models and activities so that practitioners can try different ways of solving a problem. By having reference imagery the models are guided more towards the reference and away from its base training, same with using individual styles. In our experience having generated over 100,000 images with local AI, we also found that the different models and techniques have different 'qwerks'. For instance, Stable Diffusion may have issues rendering fingers, but its inability to fully follow the prompt enables blending multiple styles. With FLUX Redux, reference images have features extracted that lead to unexpected recombinations. FlowEdit, used in Trace, adds nuance of how much distortion to create from the starting prompt to the ending prompt. Models like Wan are better at prompt adherence and consistency, but can often accidentally infringe on intellectual property if the prompt is vague.}

\revision{
One way in which design fixation can come up is when someone engages with more iterative refinement over exploration. During the study, P1 often talked about following a certain path and continuing to refine and feeling distracted: \textit{"it can be distracting because it can generate so much stuff that then you're like, oh, is this what I wanted? Or is this just exciting?}". With more physical activities, a clear limit is fatigue, but with AI generations that can go unnoticed for a long time.}

\subsection{Local Models and Implications}
\revision{
A key design choice in Atelier was to rely on local, open-source models rather than cloud services. While cloud services offer many benefits, such as not having to rely on individual hardware, much of the controllability and nuance that makes Protosampling possible is due to open source technologies. Open source models allow access all the necessary parameters to control them, making it possible to mix and match different platforms and extensions, such as the ability to add details or negative prompts or trigger multiple styles.
For example, a Paint easel can generate an image with the base model and then one can tweak it with a realism style with 30\% strength, a dream-like style with 80\% strength and a retro anime style with a 60\% style\footnote{strengths do not need to add up to 100\%, they just imply how much of the individual finetune gets embedded into the main checkpoint}. This flexibility that enables combining different models for image referencing -- with Flux Redux\cite{flux} the model automatically extracts items from each reference depending on the strength setting, whereas with USO\cite{wu2025usounifiedstylesubjectdriven} the style is explicitly isolated.}
\revision{
Other arguments for running AI locally include energy efficiency, preserving privacy and avoiding content trapped in walled gardens, and the ability to choose which models to work with. In this version of the system we preselected style models, but one can imagine the ability to import one's own finetune of a specific character or style. This brings a whole realm of possibilities for creators to truly harness the power of generative AI for creation.}

\revision{
Model flexibility is also where ethical concerns come into play. AI models were trained using publicly available data without opt-in or opt-out options, and the economic impact to the creative community has been noticeable. The ethical implications are complex and still under active discussion, since there is nothing stopping bad actors from creating malicious or inappropriate content or training on the work of specific styles when working locally. }

\subsection{Limitations}
The current design and implementation of Atelier has three main limitations. One limitation is technical. While ComfyUI is a very powerful tool for prototyping, it is not designed for robustness. Paths and models are often hard-coded, there are memory leaks, and because so much is driven by community contributions, it means that updates can break the system. Moreover, we are running the system locally on consumer hardware. While speeds can be comparable to commercial-grade hardware, it is not possible to hold all models in memory. This means that models are constantly being swapped out, which can be time consuming and adds a lot of redundancy. These limitations are not inherent to Atelier's interaction model, and as model-serving infrastructures mature, the same workflows could be deployed with significantly greater robustness and usability.

The second limitation is that Atelier runs on the assumption that the bulk of the work is done in the system. While one can import and export much of the media, the reality is that Atelier also creates another potential for information getting fragmented across applications. This is where a system like Atelier might work better if the metaphor could be applied to the operating system as a whole, similar to how it is done in other systems \cite{surfacefleet2020brudy}. \revision{That said, study participants (P2, P4) described wanting to stay in the tool, and wanting to avoid switching applications even for things like image editing for which other tools might be best suited.}

The last limitation of Atelier is conceptual. While we argue for references and generated assets to co-exist in the same virtual space, the reality is that success with generative AI requires a large number of generations \cite{ledo2025generative}. This means that a large portion of what gets generated is effectively scrap material. The other problem with having so many generations is that it is easy to lose sight of what one wants to create, whether it is because of compromise on a model's limitations, or simply because having so much content created saturates the mind and forces incidental emergence. The models also have a defined set of biases in their data and training, and those can also get perpetuated. While having multimodal inputs as done in Atelier can help with this, there is still a risk that solutions get pushed towards these biases, potentially leading to design fixation, or accidental infringement on other intellectual property. These are simply realities of dealing with AI as a medium, and will require more time and studies to better understand.

\subsection{Future Work}
The work carried out in Atelier opens a set of interesting research opportunities. The first one is the exploration of video, given that thus far we only explored image-to-video. There are many more ways to create generative video that have even more controllability \cite{ledo2025generative}, and those modalities need appropriate abstractions and gaining a better understanding of the interactions at play. The canvas approach for video invites thinking about abstractions, but also reflects on how there might be potential interplays between the physical space and the time-based nature of animated motion.

Another potential area is exploring how to make more technically robust infrastructures accessible. For example, Atelier currently supports a small set of LoRAs and checkpoints, yet being able to have specifically trained characters could dramatically improve the different activities supported by the easels. There are also questions for what it means to add more intelligence to the canvas, or support chaining or nesting workflows in a way that does not require computational thinking.

Lastly, exploring protosampling and operationalizing it invites research into having a better understanding of how practitioners might use Atelier for their own creative process, and seeing how their ideas and thoughts evolve over time. Interesting future work could involve practitioners using the system over a period of time to see what they can create, what strategies they devise, and how different people make use of the canvas and the tools provided.

\section{Conclusion}
Sawyer's compilation of creative theories show how insight, no matter how sudden it may seem, is a systematic process always connected to a practitioner's prior work, their sketches, notebooks, etc. \cite{sawyer2024explaining}. This highlights the importance of curating collections, manifesting ideas, and reflecting on the process. Tightly interweaving sampling and prototyping -- \textit{Protosampling}-- helps better understand the relationship between the collected information and how ideas get manifested. This is especially true now that generative AI has accelerated and blurred the boundaries between these two important activities. Deriving and operationalizing Protosampling, and understanding how information moves about in the process enabled designing Atelier as one open-ended way to blend thinking and making, abstract complex workflows into easels as self-contained modules, and make sense of and organize the work.

Atelier was built with Protosampling at its core. Atelier looks to make the creative process more tangible, and speaks to how creative work in this new age of generative AI is not about finding the right prompt, or accessing the latest model, but by the small decisions and the deliberate actions creators take, and how the right tools can empower them to make their ideas come true.


\balance
\bibliographystyle{ACM-Reference-Format}
\bibliography{refs}

@inproceedings{misty,
author = {Lu, Yuwen and Leung, Alan and Swearngin, Amanda and Nichols, Jeffrey and Barik, Titus},
title = {Misty: UI Prototyping Through Interactive Conceptual Blending},
year = {2025},
isbn = {9798400713941},
publisher = {Association for Computing Machinery},
address = {New York, NY, USA},
url = {https://doi-org.offcampus.lib.washington.edu/10.1145/3706598.3713924},
doi = {10.1145/3706598.3713924},
booktitle = {Proceedings of the 2025 CHI Conference on Human Factors in Computing Systems},
articleno = {1108},
numpages = {17},
location = {
},
series = {CHI '25}
}

@incollection{kochanowska2021double,
  title={The double diamond model: In pursuit of simplicity and flexibility},
  author={Kochanowska, Magda and Gagliardi, Weronika Rochacka},
  booktitle={Perspectives on design II: Research, education and practice},
  pages={19--32},
  year={2021},
  publisher={Springer}
}

@inproceedings{creativeconnect,
author = {Choi, DaEun and Hong, Sumin and Park, Jeongeon and Chung, John Joon Young and Kim, Juho},
title = {CreativeConnect: Supporting Reference Recombination for Graphic Design Ideation with Generative AI},
year = {2024},
isbn = {9798400703300},
publisher = {Association for Computing Machinery},
address = {New York, NY, USA},
url = {https://doi-org.offcampus.lib.washington.edu/10.1145/3613904.3642794},
doi = {10.1145/3613904.3642794},
booktitle = {Proceedings of the 2024 CHI Conference on Human Factors in Computing Systems},
articleno = {1055},
numpages = {25},
location = {Honolulu, HI, USA},
series = {CHI '24}
}

@inproceedings{neuralcanvas,
author = {Shen, Yulin and Shen, Yifei and Cheng, Jiawen and Jiang, Chutian and Fan, Mingming and Wang, Zeyu},
title = {Neural Canvas: Supporting Scenic Design Prototyping by Integrating 3D Sketching and Generative AI},
year = {2024},
isbn = {9798400703300},
publisher = {Association for Computing Machinery},
address = {New York, NY, USA},
url = {https://doi-org.offcampus.lib.washington.edu/10.1145/3613904.3642096},
doi = {10.1145/3613904.3642096},
articleno = {1056},
numpages = {18},
location = {Honolulu, HI, USA},
series = {CHI '24}
}

@inproceedings{vrcopilot,
author = {Zhang, Lei and Pan, Jin and Gettig, Jacob and Oney, Steve and Guo, Anhong},
title = {VRCopilot: Authoring 3D Layouts with Generative AI Models in VR},
year = {2024},
isbn = {9798400706288},
publisher = {Association for Computing Machinery},
address = {New York, NY, USA},
url = {https://doi-org.offcampus.lib.washington.edu/10.1145/3654777.3676451},
doi = {10.1145/3654777.3676451},
booktitle = {Proceedings of the 37th Annual ACM Symposium on User Interface Software and Technology},
articleno = {96},
numpages = {13},
location = {Pittsburgh, PA, USA},
series = {UIST '24}
}

@online{sculpt,
author = {Jukka Seppänen},
title= {ComfyUI-Hunyuan3DWrapper},
url = {https://github.com/kijai/ComfyUI-Hunyuan3DWrapper},
urldate={2025-09-11},
Year={2025}}

@article{bostock2011d3,
  title={D$^3$ data-driven documents},
  author={Bostock, Michael and Ogievetsky, Vadim and Heer, Jeffrey},
  journal={IEEE transactions on visualization and computer graphics},
  volume={17},
  number={12},
  pages={2301--2309},
  year={2011},
  publisher={IEEE}
}

@article{walny2016thinking,
  title={Thinking with Sketches: Leveraging Everyday Use of Visuals for Information Visualization},
  author={Walny, Jagoda},
  year={2016}
}

@inproceedings{marquardt2025imaginationvellum,
  title={ImaginationVellum: Generative-AI Ideation Canvas with Spatial Prompts, Generative Strokes, and Ideation History},
  author={Marquardt, Nicolai and Roseway, Asta and Romat, Hugo and Panda, Payod and Pahud, Michel and Ramos, Gonzalo and Drucker, Steven M and Wilson, Andrew D and Hinckley, Ken and Riche, Nathalie},
  booktitle={Proceedings of the 38th Annual ACM Symposium on User Interface Software and Technology},
  pages={1--19},
  year={2025}
}

@article{visualstories2024antony,
author = {Antony, Victor Nikhil and Huang, Chien-Ming},
title = {ID.8: Co-Creating Visual Stories with Generative AI},
year = {2024},
issue_date = {September 2024},
publisher = {Association for Computing Machinery},
address = {New York, NY, USA},
volume = {14},
number = {3},
issn = {2160-6455},
url = {https://doi-org.offcampus.lib.washington.edu/10.1145/3672277},
doi = {10.1145/3672277},
journal = {ACM Trans. Interact. Intell. Syst.},
month = aug,
articleno = {20},
numpages = {29}
}

@inproceedings{surfacefleet2020brudy,
author = {Brudy, Frederik and Ledo, David and Pahud, Michel and Henry Riche, Nathalie and Holz, Christian and Waghmare, Anand and Surale, Hemant Bhaskar and Peinado, Marcus and Zhang, Xiaokuan and Joyner, Shannon and Chandramouli, Badrish and Minhas, Umar Farooq and Goldstein, Jonathan and Buxton, William and Hinckley, Ken},
title = {SurfaceFleet: Exploring Distributed Interactions Unbounded from Device, Application, User, and Time},
year = {2020},
isbn = {9781450375146},
publisher = {Association for Computing Machinery},
address = {New York, NY, USA},
url = {https://doi-org.offcampus.lib.washington.edu/10.1145/3379337.3415874},
doi = {10.1145/3379337.3415874},
booktitle = {Proceedings of the 33rd Annual ACM Symposium on User Interface Software and Technology},
pages = {7–21},
numpages = {15},
location = {Virtual Event, USA},
series = {UIST '20}
}

@article{realfill2024tang,
author = {Tang, Luming and Ruiz, Nataniel and Chu, Qinghao and Li, Yuanzhen and Holynski, Aleksander and Jacobs, David E. and Hariharan, Bharath and Pritch, Yael and Wadhwa, Neal and Aberman, Kfir and Rubinstein, Michael},
title = {RealFill: Reference-Driven Generation for Authentic Image Completion},
year = {2024},
issue_date = {July 2024},
publisher = {Association for Computing Machinery},
address = {New York, NY, USA},
volume = {43},
number = {4},
issn = {0730-0301},
url = {https://doi-org.offcampus.lib.washington.edu/10.1145/3658237},
doi = {10.1145/3658237},
journal = {ACM Trans. Graph.},
month = jul,
articleno = {135},
numpages = {12}
}

@inproceedings{composable2025amin,
author = {Amin, Rifat Mehreen and K\"{u}hle, Oliver Hans and Buschek, Daniel and Butz, Andreas},
title = {Composable Prompting Workspaces for Creative Writing: Exploration and Iteration Using Dynamic Widgets},
year = {2025},
isbn = {9798400713958},
publisher = {Association for Computing Machinery},
address = {New York, NY, USA},
url = {https://doi-org.offcampus.lib.washington.edu/10.1145/3706599.3720243},
doi = {10.1145/3706599.3720243},
booktitle = {Proceedings of the Extended Abstracts of the CHI Conference on Human Factors in Computing Systems},
articleno = {144},
numpages = {11},
location = {
},
series = {CHI EA '25}
}

@inproceedings{designsketches2023zhang,
author = {Zhang, Chengzhi and Wang, Weijie and Pangaro, Paul and Martelaro, Nikolas and Byrne, Daragh},
title = {Generative Image AI Using Design Sketches as input: Opportunities and Challenges},
year = {2023},
isbn = {9798400701801},
publisher = {Association for Computing Machinery},
address = {New York, NY, USA},
url = {https://doi-org.offcampus.lib.washington.edu/10.1145/3591196.3596820},
doi = {10.1145/3591196.3596820},
booktitle = {Proceedings of the 15th Conference on Creativity and Cognition},
pages = {254–261},
numpages = {8},
location = {Virtual Event, USA},
series = {C\&C '23}
}

@inproceedings{scaffoldingsketch2024,
author = {Sarukkai, Vishnu and Yuan, Lu and Tang, Mia and Agrawala, Maneesh and Fatahalian, Kayvon},
title = {Block and Detail: Scaffolding Sketch-to-Image Generation},
year = {2024},
isbn = {9798400706288},
publisher = {Association for Computing Machinery},
address = {New York, NY, USA},
url = {https://doi-org.offcampus.lib.washington.edu/10.1145/3654777.3676444},
doi = {10.1145/3654777.3676444},
booktitle = {Proceedings of the 37th Annual ACM Symposium on User Interface Software and Technology},
articleno = {33},
numpages = {13},
location = {Pittsburgh, PA, USA},
series = {UIST '24}
}

@inproceedings{promptpaint2023chung,
author = {Chung, John Joon Young and Adar, Eytan},
title = {PromptPaint: Steering Text-to-Image Generation Through Paint Medium-like Interactions},
year = {2023},
isbn = {9798400701320},
publisher = {Association for Computing Machinery},
address = {New York, NY, USA},
url = {https://doi-org.offcampus.lib.washington.edu/10.1145/3586183.3606777},
doi = {10.1145/3586183.3606777},
booktitle = {Proceedings of the 36th Annual ACM Symposium on User Interface Software and Technology},
articleno = {6},
numpages = {17},
location = {San Francisco, CA, USA},
series = {UIST '23}
}

@inproceedings{sketchflex2025lin,
author = {Lin, Haichuan and Ye, Yilin and Xia, Jiazhi and Zeng, Wei},
title = {SketchFlex: Facilitating Spatial-Semantic Coherence in Text-to-Image Generation with Region-Based Sketches},
year = {2025},
isbn = {9798400713941},
publisher = {Association for Computing Machinery},
address = {New York, NY, USA},
url = {https://doi-org.offcampus.lib.washington.edu/10.1145/3706598.3713801},
doi = {10.1145/3706598.3713801},
booktitle = {Proceedings of the 2025 CHI Conference on Human Factors in Computing Systems},
articleno = {546},
numpages = {19},
location = {
},
series = {CHI '25}
}

@inproceedings{aideation2025wang,
author = {Wang, Wen-Fan and Lu, Chien-Ting and Ponsa i Campany\`{a}, Nil and Chen, Bing-Yu and Chen, Mike Y.},
title = {AIdeation: Designing a Human-AI Collaborative Ideation System for Concept Designers},
year = {2025},
isbn = {9798400713941},
publisher = {Association for Computing Machinery},
address = {New York, NY, USA},
url = {https://doi-org.offcampus.lib.washington.edu/10.1145/3706598.3714148},
doi = {10.1145/3706598.3714148},
booktitle = {Proceedings of the 2025 CHI Conference on Human Factors in Computing Systems},
articleno = {21},
numpages = {28},
location = {
},
series = {CHI '25}
}

@inproceedings{lucero2012moodboards,
author = {Lucero, Andr\'{e}s},
title = {Framing, aligning, paradoxing, abstracting, and directing: how design mood boards work},
year = {2012},
isbn = {9781450312103},
publisher = {Association for Computing Machinery},
address = {New York, NY, USA},
url = {https://doi-org.offcampus.lib.washington.edu/10.1145/2317956.2318021},
doi = {10.1145/2317956.2318021},
booktitle = {Proceedings of the Designing Interactive Systems Conference},
pages = {438–447},
numpages = {10},
location = {Newcastle Upon Tyne, United Kingdom},
series = {DIS '12}
}

@article{moodboardsdesign2004mcdonagh,
author = {Deana Mcdonagh and Ian Storer},
title = {Mood Boards as a Design Catalyst and Resource: Researching an Under-Researched Area},
journal = {The Design Journal},
volume = {7},
number = {3},
pages = {16--31},
year = {2004},
publisher = {Routledge},
doi = {10.2752/146069204789338424},
URL = { https://doi.org/10.2752/146069204789338424
},
eprint = { https://doi.org/10.2752/146069204789338424
}
}

@inproceedings{gancollage2023wan,
author = {Wan, Qian and Lu, Zhicong},
title = {GANCollage: A GAN-Driven Digital Mood Board to Facilitate Ideation in Creativity Support},
year = {2023},
isbn = {9781450398930},
publisher = {Association for Computing Machinery},
address = {New York, NY, USA},
url = {https://doi-org.offcampus.lib.washington.edu/10.1145/3563657.3596072},
doi = {10.1145/3563657.3596072},
booktitle = {Proceedings of the 2023 ACM Designing Interactive Systems Conference},
pages = {136–146},
numpages = {11},
location = {Pittsburgh, PA, USA},
series = {DIS '23}
}

@inproceedings{neustaedter2012autobiographical,
  title={Autobiographical design in HCI research: designing and learning through use-it-yourself},
  author={Neustaedter, Carman and Sengers, Phoebe},
  booktitle={Proceedings of the Designing Interactive Systems Conference},
  pages={514--523},
  year={2012}
}

@inproceedings{peng2024designprompt,
  title={Designprompt: Using multimodal interaction for design exploration with generative ai},
  author={Peng, Xiaohan and Koch, Janin and Mackay, Wendy E},
  booktitle={Proceedings of the 2024 ACM Designing Interactive Systems Conference},
  pages={804--818},
  year={2024}
}

@inproceedings{leong2025paratrouper,
  title={Paratrouper: Exploratory Creation of Character Cast Visuals Using Generative AI},
  author={Leong, Joanne and Ledo, David and Driscoll, Thomas and Grossman, Tovi and Fitzmaurice, George and Anderson, Fraser},
  booktitle={Proceedings of the 2025 CHI Conference on Human Factors in Computing Systems},
  pages={1--20},
  year={2025}
}

@inproceedings{ledo2025generative,
  title={Generative Rotoscoping: A First-Person Autobiographical Exploration on Generative Video-to-Video Practices},
  author={Ledo, David},
  booktitle={Proceedings of the 2025 Conference on Creativity and Cognition},
  pages={931--948},
  year={2025}
}

@article{ho2022classifier,
  title={Classifier-free diffusion guidance},
  author={Ho, Jonathan and Salimans, Tim},
  journal={arXiv preprint arXiv:2207.12598},
  year={2022}
}

@article{chen2025normalized,
  title={Normalized Attention Guidance: Universal Negative Guidance for Diffusion Model},
  author={Chen, Dar-Yen and Bandyopadhyay, Hmrishav and Zou, Kai and Song, Yi-Zhe},
  journal={arXiv preprint arXiv:2505.21179},
  year={2025}
}

@misc{wu2025usounifiedstylesubjectdriven,
      title={USO: Unified Style and Subject-Driven Generation via Disentangled and Reward Learning}, 
      author={Shaojin Wu and Mengqi Huang and Yufeng Cheng and Wenxu Wu and Jiahe Tian and Yiming Luo and Fei Ding and Qian He},
      year={2025},
      eprint={2508.18966},
      archivePrefix={arXiv},
      primaryClass={cs.CV},
      url={https://arxiv.org/abs/2508.18966}, 
}

@article{yang2024depth,
  title={Depth anything v2},
  author={Yang, Lihe and Kang, Bingyi and Huang, Zilong and Zhao, Zhen and Xu, Xiaogang and Feng, Jiashi and Zhao, Hengshuang},
  journal={Advances in Neural Information Processing Systems},
  volume={37},
  pages={21875--21911},
  year={2024}
}

@inproceedings{xiao2024florence,
  title={Florence-2: Advancing a unified representation for a variety of vision tasks},
  author={Xiao, Bin and Wu, Haiping and Xu, Weijian and Dai, Xiyang and Hu, Houdong and Lu, Yumao and Zeng, Michael and Liu, Ce and Yuan, Lu},
  booktitle={Proceedings of the IEEE/CVF Conference on Computer Vision and Pattern Recognition},
  pages={4818--4829},
  year={2024}
}

@misc{wan2025wanopenadvancedlargescale,
      title={Wan: Open and Advanced Large-Scale Video Generative Models}, 
      author={Team Wan and Ang Wang and Baole Ai and Bin Wen and Chaojie Mao and Chen-Wei Xie and Di Chen and Feiwu Yu and Haiming Zhao and Jianxiao Yang and Jianyuan Zeng and Jiayu Wang and Jingfeng Zhang and Jingren Zhou and Jinkai Wang and Jixuan Chen and Kai Zhu and Kang Zhao and Keyu Yan and Lianghua Huang and Mengyang Feng and Ningyi Zhang and Pandeng Li and Pingyu Wu and Ruihang Chu and Ruili Feng and Shiwei Zhang and Siyang Sun and Tao Fang and Tianxing Wang and Tianyi Gui and Tingyu Weng and Tong Shen and Wei Lin and Wei Wang and Wei Wang and Wenmeng Zhou and Wente Wang and Wenting Shen and Wenyuan Yu and Xianzhong Shi and Xiaoming Huang and Xin Xu and Yan Kou and Yangyu Lv and Yifei Li and Yijing Liu and Yiming Wang and Yingya Zhang and Yitong Huang and Yong Li and You Wu and Yu Liu and Yulin Pan and Yun Zheng and Yuntao Hong and Yupeng Shi and Yutong Feng and Zeyinzi Jiang and Zhen Han and Zhi-Fan Wu and Ziyu Liu},
      year={2025},
      eprint={2503.20314},
      archivePrefix={arXiv},
      primaryClass={cs.CV},
      url={https://arxiv.org/abs/2503.20314}, 
}

@online{RunwayML,
  author = {RunwayML},
  title = {Gen-3 Alpha},
  year = 2024,
  url = {https://runwayml.com/},
  urldate = {February 4, 2025}
}

@online{FigmaWeave,
  author = {Figma},
  title = {Introducing Figma Weave: The next generation of AI-native creation at Figma},
  year = 2025,
  url = {https://www.figma.com/blog/welcome-weavy-to-figma/},
  urldate = {December 4, 2025}
}

@online{ComfyUI,
  author = {ComfyUI},
  year = 2024,
  url = {https://www.comfy.org/},
  urldate = {February 4, 2025}
}

@online{FluxTurbo,
  author = {Alibaba},
  title = {Flux Turbo},
  year = 2024,
  url = {https://huggingface.co/alimama-creative/FLUX.1-Turbo-Alpha},
  urldate = {February 4, 2025}
}

@phdthesis{martinez2019openpose,
  title={Openpose: Whole-body pose estimation},
  author={Mart{\i}nez, Gin{\'e}s Hidalgo},
  year={2019},
  school={Carnegie Mellon University Pittsburgh, PA, USA}
}

@article{kulikov2024flowedit,
  title={FlowEdit: Inversion-Free Text-Based Editing Using Pre-Trained Flow Models},
  author={Kulikov, Vladimir and Kleiner, Matan and Huberman-Spiegelglas, Inbar and Michaeli, Tomer},
  journal={arXiv preprint arXiv:2412.08629},
  year={2024}
}

@inproceedings{dix2011externalisation,
  title={Externalisation and design},
  author={Dix, Alan and Gongora, Layda},
  booktitle={Procedings of the second conference on creativity and innovation in design},
  pages={31--42},
  year={2011}
}

@book{buxton2010sketching,
  title={Sketching user experiences: getting the design right and the right design},
  author={Buxton, Bill},
  year={2010},
  publisher={Morgan kaufmann}
}

@inproceedings{moodcubesIvanov,
author = {Ivanov, Alexander and Ledo, David and Grossman, Tovi and Fitzmaurice, George and Anderson, Fraser},
title = {MoodCubes: Immersive Spaces for Collecting, Discovering and Envisioning Inspiration Materials},
year = {2022},
isbn = {9781450393584},
publisher = {Association for Computing Machinery},
address = {New York, NY, USA},
url = {https://doi.org/10.1145/3532106.3533565},
doi = {10.1145/3532106.3533565},
booktitle = {Proceedings of the 2022 ACM Designing Interactive Systems Conference},
pages = {189–203},
numpages = {15},
location = {Virtual Event, Australia},
series = {DIS '22}
}

@inproceedings{zhang2023controlnet,
  title={Adding conditional control to text-to-image diffusion models},
  author={Zhang, Lvmin and Rao, Anyi and Agrawala, Maneesh},
  booktitle={Proceedings of the IEEE/CVF International Conference on Computer Vision},
  pages={3836--3847},
  year={2023}
}

@article{zhao2024uni,
  title={Uni-controlnet: All-in-one control to text-to-image diffusion models},
  author={Zhao, Shihao and Chen, Dongdong and Chen, Yen-Chun and Bao, Jianmin and Hao, Shaozhe and Yuan, Lu and Wong, Kwan-Yee K},
  journal={Advances in Neural Information Processing Systems},
  volume={36},
  year={2024}
}

@article{kochImageSense,
author = {Koch, Janin and Taffin, Nicolas and Beaudouin-Lafon, Michel and Laine, Markku and Lucero, Andr\'{e}s and Mackay, Wendy E.},
title = {ImageSense: An Intelligent Collaborative Ideation Tool to Support Diverse Human-Computer Partnerships},
year = {2020},
issue_date = {May 2020},
publisher = {Association for Computing Machinery},
address = {New York, NY, USA},
volume = {4},
number = {CSCW1},
url = {https://doi.org/10.1145/3392850},
doi = {10.1145/3392850},
journal = {Proc. ACM Hum.-Comput. Interact.},
month = {may},
articleno = {45},
numpages = {27}
}

@inproceedings{hinckleyInkseine,
author = {Hinckley, Ken and Zhao, Shengdong and Sarin, Raman and Baudisch, Patrick and Cutrell, Edward and Shilman, Michael and Tan, Desney},
title = {InkSeine: In Situ search for active note taking},
year = {2007},
isbn = {9781595935939},
publisher = {Association for Computing Machinery},
address = {New York, NY, USA},
url = {https://doi.org/10.1145/1240624.1240666},
doi = {10.1145/1240624.1240666},
booktitle = {Proceedings of the SIGCHI Conference on Human Factors in Computing Systems},
pages = {251–260},
numpages = {10},
location = {San Jose, California, USA},
series = {CHI '07}
}

@inproceedings{xia2016objectorienteddrawing,
  title={Object-oriented drawing},
  author={Xia, Haijun and Araujo, Bruno and Grossman, Tovi and Wigdor, Daniel},
  booktitle={Proceedings of the 2016 CHI Conference on Human Factors in Computing Systems},
  pages={4610--4621},
  year={2016}
}

@inproceedings{wiberg2014makes,
  title={What makes a prototype novel? A knowledge contribution concern for interaction design research},
  author={Wiberg, Mikael and Stolterman, Erik},
  booktitle={Proceedings of the 8th Nordic conference on human-computer interaction: fun, fast, foundational},
  pages={531--540},
  year={2014}
}

@article{ohlsson1992information,
  title={Information-processing explanations of insight and related phenomena},
  author={Ohlsson, Stellan},
  journal={Advances in the psychology of thinking},
  pages={1--44},
  year={1992},
  publisher={Harvester Wheatsheaf}
}

@article{weisberg1993creativity,
  title={Creativity: Beyond the myth of genius},
  author={Weisberg, Robert W},
  journal={(No Title)},
  year={1993}
}

@book{weisberg1986creativity,
  title={Creativity: Genius and other myths.},
  author={Weisberg, Robert},
  year={1986},
  publisher={WH Freeman/Times Books/Henry Holt \& Co}
}

@inproceedings{gaver2022emergence,
  title={Emergence as a feature of practice-based design research},
  author={Gaver, William and Krogh, Peter Gall and Boucher, Andy and Chatting, David},
  booktitle={Proceedings of the 2022 ACM designing interactive systems conference},
  pages={517--526},
  year={2022}
}

@article{goel1992structure,
  title={The structure of design problem spaces},
  author={Goel, Vinod and Pirolli, Peter},
  journal={Cognitive science},
  volume={16},
  number={3},
  pages={395--429},
  year={1992},
  publisher={Elsevier}
}

@article{ortony1979beyond,
  title={Beyond literal similarity.},
  author={Ortony, Andrew},
  journal={Psychological review},
  volume={86},
  number={3},
  pages={161},
  year={1979},
  publisher={American Psychological Association}
}

@article{makel2014creativity,
  title={Creativity is more than novelty: Reconsidering replication as a creativity act.},
  author={Makel, Matthew C and Plucker, Jonathan A},
  year={2014},
  publisher={Educational Publishing Foundation}
}

@article{lockhart1988conceptual,
  title={Conceptual transfer insimple insight problems},
  author={Lockhart, Robert S and Lamon, Mary and Gick, Mary L},
  journal={Memory \& Cognition},
  volume={16},
  number={1},
  pages={36--44},
  year={1988},
  publisher={Springer}
}

@article{mobley1992process,
  title={Process analytic models of creative capacities: Evidence for the combination and reorganization process},
  author={Mobley, Michele I and Doares, Lesli M and Mumford, Michael D},
  journal={Creativity Research Journal},
  volume={5},
  number={2},
  pages={125--155},
  year={1992},
  publisher={Taylor \& Francis}
}

@book{sellen2003myth,
  title={The myth of the paperless office},
  author={Sellen, Abigail J and Harper, Richard HR},
  year={2003},
  publisher={MIT press}
}

@book{blandford2016qualitativehci,
  title={Qualitative HCI research: Going behind the scenes},
  author={Blandford, Ann and Furniss, Dominic and Makri, Stephann},
  year={2016},
  publisher={Morgan \& Claypool Publishers}
}

@inproceedings{dang2023worldsmith,
  title={WorldSmith: Iterative and Expressive Prompting for World Building with a Generative AI},
  author={Dang, Hai and Brudy, Frederik and Fitzmaurice, George and Anderson, Fraser},
  booktitle={Proceedings of the 36th Annual ACM Symposium on User Interface Software and Technology},
  pages={1--17},
  year={2023}
}

@article{sawyer2021iterativecreativeprocess,
  title={The iterative and improvisational nature of the creative process},
  author={Sawyer, R Keith},
  journal={Journal of Creativity},
  volume={31},
  pages={100002},
  year={2021},
  publisher={Elsevier}
}

@article{lim2008anatomyofprototypes,
  title={The anatomy of prototypes: Prototypes as filters, prototypes as manifestations of design ideas},
  author={Lim, Youn-Kyung and Stolterman, Erik and Tenenberg, Josh},
  journal={ACM Transactions on Computer-Human Interaction (TOCHI)},
  volume={15},
  number={2},
  pages={1--27},
  year={2008},
  publisher={ACM New York, NY, USA}
}

@inproceedings{stemasov2023immersivesampling,
  title={Immersive Sampling: Exploring Sampling for Future Creative Practices in Media-Rich, Immersive Spaces},
  author={Stemasov, Evgeny and Ledo, David and Fitzmaurice, George and Anderson, Fraser},
  booktitle={Proceedings of the 2023 ACM Designing Interactive Systems Conference},
  pages={212--229},
  year={2023}
}

@article{eckert2000sources,
  title={Sources of inspiration: a language of design},
  author={Eckert, Claudia and Stacey, Martin},
  journal={Design studies},
  volume={21},
  number={5},
  pages={523--538},
  year={2000},
  publisher={Elsevier}
}

@online{chatgpt,
author = {OpenAI},
title = "ChatGPT",
url = "https://chatgpt.com/overview/",
urldate={2025-09-11},
Year={2025}}

@online{Cursor,
author = {Anysphere},
title = "Cursor",
url = {https://cursor.com/home},
urldate={2025-09-11},
Year={2025}}

@online{tldraw,
author = {tldraw},
title = "tldraw",
url = "https://tldraw.dev/quick-start",
urldate={2025-09-11},
Year={2025}}

@online{FLORA,
author = {FLORA},
title = "FLORA",
url = "https://www.florafauna.ai/academy/quickstart",
urldate={2025-09-11},
Year={2025}}

@article{smith1984conceptual,
  title={Conceptual combination with prototype concepts},
  author={Smith, Edward E and Osherson, Daniel N},
  journal={Cognitive science},
  volume={8},
  number={4},
  pages={337--361},
  year={1984},
  publisher={Wiley Online Library}
}

@online{Mural,
author = {Mural},
title = "FLORA",
url = "https://www.mural.co/use-case/design-canvas",
urldate={2025-12-04},
Year={2025}}

@online{Miro,
author = {Miro},
title = "Miro",
url = "https://miro.com/",
urldate={2025-12-04},
Year={2025}}

@incollection{floyd1984systematic,
  title={A systematic look at prototyping},
  author={Floyd, Christiane},
  booktitle={Approaches to prototyping},
  pages={1--18},
  year={1984},
  publisher={Springer}
}

@online{Milanote,
author = {Milanote},
title = "Milanote",
url = "https://milanote.com/",
urldate={2025-12-04},
Year={2025}}

@online{Adobefirefly,
author = {Adobe},
title = "Adobe Firefly",
url = "https://firefly.adobe.com/",
urldate={2025-12-04},
Year={2025}}

@online{Dubberly,
title = "How do you design?",
url = "https://www.dubberly.com/articles/how-do-you-design.html",
urldate = {2025-12-04},
year = {2005}}

@book{boden2004creative,
  title={The creative mind: Myths and mechanisms},
  author={Boden, Margaret A},
  year={2004},
  publisher={Routledge}
}

@online{dalle,
    author = {OpenAI},
    title = "DALLE·3",
    url = {https://openai.com/index/dall-e-3/},
    urldate = {2024-11-10},
    year = {2024},
}

@online{midjourney,
    author = {Midjourney},
    title = "Midjourney Documentation",
    url = {https://docs.midjourney.com/},
    urldate = {2024-11-10},
    year = {2024},
}

@book{dourish2001action,
  title={Where the action is: the foundations of embodied interaction},
  author={Dourish, Paul},
  year={2001},
  publisher={MIT press}
}

@incollection{logan2013creativity,
  title={Creativity and design as exploration},
  author={Logan, Brian and Smithers, Tim},
  booktitle={Modeling creativity and knowledge-based creative design},
  pages={139--175},
  year={2013},
  publisher={Psychology Press}
}

@inproceedings{bondarenko2005documents,
  title={Documents at hand: Learning from paper to improve digital technologies},
  author={Bondarenko, Olha and Janssen, Ruud},
  booktitle={Proceedings of the SIGCHI conference on Human factors in computing systems},
  pages={121--130},
  year={2005}
}

@article{oren_overview_2006,
	title = {An {Overview} of {Information} {Management} and {Knowledge} {Work} {Studies}: {Lessons} for the {Semantic} {Desktop}.},
	shorttitle = {An {Overview} of {Information} {Management} and {Knowledge} {Work} {Studies}},
	journal = {SemDesk},
	author = {Oren, Eyal},
	year = {2006},
	note = {Publisher: Citeseer},
}

@book{jones2010keeping,
  title={Keeping found things found: The study and practice of personal information management},
  author={Jones, William},
  year={2010},
  publisher={Morgan Kaufmann}
}

@article{malone1983people,
  title={How do people organize their desks? Implications for the design of office information systems},
  author={Malone, Thomas W},
  journal={ACM Transactions on Information Systems (TOIS)},
  volume={1},
  number={1},
  pages={99--112},
  year={1983},
  publisher={ACM New York, NY, USA}
}

@inproceedings{henderson2009empirical,
  title={An empirical analysis of personal digital document structures},
  author={Henderson, Sarah and Srinivasan, Ananth},
  booktitle={Symposium on Human Interface},
  pages={394--403},
  year={2009},
  organization={Springer}
}

@article{oren2006overview,
  title={An Overview of Information Management and Knowledge Work Studies: Lessons for the Semantic Desktop.},
  author={Oren, Eyal},
  journal={SemDesk},
  year={2006}
}

@article{bondarenko2010requirements,
  title={Requirements for the design of a personal document-management system},
  author={Bondarenko, Olha and Janssen, Ruud and Driessen, Samu{\"e}l},
  journal={Journal of the American Society for Information Science and Technology},
  volume={61},
  number={3},
  pages={468--482},
  year={2010},
  publisher={Wiley Online Library}
}

@inproceedings{kidd1994marks,
  title={The marks are on the knowledge worker},
  author={Kidd, Alison},
  booktitle={Proceedings of the SIGCHI conference on Human factors in computing systems},
  pages={186--191},
  year={1994}
}

@article{christie2014language,
  title={Language helps children succeed on a classic analogy task},
  author={Christie, Stella and Gentner, Dedre},
  journal={Cognitive science},
  volume={38},
  number={2},
  pages={383--397},
  year={2014},
  publisher={Wiley Online Library}
}

@incollection{gero1994computational,
  title={Computational models of creative design processes},
  author={Gero, John S},
  booktitle={Artificial intelligence and creativity: An interdisciplinary approach},
  pages={269--281},
  year={1994},
  publisher={Springer}
}

@article{horng2021behavioural,
  title={A behavioural measure of imagination based on conceptual combination theory},
  author={Horng, Ruey-Yun and Wang, Ching-Wen and Yen, Yung-Chieh and Lu, Chia-Ying and Li, Chien-Tao},
  journal={Creativity Research Journal},
  volume={33},
  number={4},
  pages={376--387},
  year={2021},
  publisher={Taylor \& Francis}
}

@article{sternberg2003development,
  title={The development of creativity as a decision-making process},
  author={Sternberg, Robert J},
  journal={Creativity and development},
  volume={1},
  pages={91--138},
  year={2003}
}

@online{flux,
    author = {FLUX.1},
    title = "FLUX.1",
    url = {https://getimg.ai/models/flux},
    urldate = {2024-11-10},
    year = {2024},
}

@online{stablediffusion,
    author = {Stability AI},
    title = "Stability AI",
    url = {https://stability.ai/},
    urldate = {2024-11-10},
    year = {2024},
}

@book{kaufman2016creativity,
  title={Creativity 101},
  author={Kaufman, James C},
  year={2016},
  publisher={Springer publishing company}
}

@book{sawyer2024explaining,
  title={Explaining creativity: The science of human innovation},
  author={Sawyer, Robert Keith and Henriksen, Danah},
  year={2024},
  publisher={Oxford university press}
}

@inproceedings{frich2019strategies,
  title={Strategies in Creative Professionals' Use of Digital Tools Across Domains},
  author={Frich, Jonas and Biskjaer, Michael Mose and MacDonald Vermeulen, Lindsay and Remy, Christian and Dalsgaard, Peter},
  booktitle={Proceedings of the 2019 Conference on Creativity and Cognition},
  pages={210--221},
  year={2019}
}

@article{cross1997descriptive,
  title={Descriptive models of creative design: application to an example},
  author={Cross, Nigel},
  journal={Design studies},
  volume={18},
  number={4},
  pages={427--440},
  year={1997},
  publisher={Elsevier}
}

@article{cross2011understanding,
  title={Understanding design thinking},
  author={Cross, Nigel},
  journal={Notes on doctoral re-search in design: Contributions from the Politecnico Di Milano. FrancoAngeli},
  pages={p19--37},
  year={2011}
}

@inproceedings{ledo2018evaluation,
  title={Evaluation strategies for HCI toolkit research},
  author={Ledo, David and Houben, Steven and Vermeulen, Jo and Marquardt, Nicolai and Oehlberg, Lora and Greenberg, Saul},
  booktitle={Proceedings of the 2018 CHI Conference on Human Factors in Computing Systems},
  pages={1--17},
  year={2018}
}

@inproceedings{olsen2007evaluating,
  title={Evaluating user interface systems research},
  author={Olsen Jr, Dan R},
  booktitle={Proceedings of the 20th annual ACM symposium on User interface software and technology},
  pages={251--258},
  year={2007}
}

@article{myers2000past,
  title={Past, present, and future of user interface software tools},
  author={Myers, Brad and Hudson, Scott E and Pausch, Randy},
  journal={ACM Transactions on Computer-Human Interaction (TOCHI)},
  volume={7},
  number={1},
  pages={3--28},
  year={2000},
  publisher={ACM New York, NY, USA}
}

@book{schon2017reflective,
  title={The reflective practitioner: How professionals think in action},
  author={Sch{\"o}n, Donald A},
  year={2017},
  publisher={Routledge}
}

@inproceedings{kochSemanticCollageEnrichingDigital2020,
  title = {{{SemanticCollage}}: {{Enriching Digital Mood Board Design}} with {{Semantic Labels}}},
  shorttitle = {{{SemanticCollage}}},
  booktitle = {{{DIS}} '20 - {{Designing Interactive Systems Conference}} 2020},
  author = {Koch, Janin and Taffin, Nicolas and Lucero, Andr{\'e}s and Mackay, Wendy E.},
  year = {2020},
  month = jul,
  pages = {407},
  publisher = {{ACM}},
  doi = {10.1145/3357236.3395494},
  urldate = {2022-05-26},
  langid = {english}
}

@inproceedings{luceroFramingAligningParadoxing2012,
  title = {Framing, Aligning, Paradoxing, Abstracting, and Directing: How Design Mood Boards Work},
  shorttitle = {Framing, Aligning, Paradoxing, Abstracting, and Directing},
  booktitle = {Proceedings of the {{Designing Interactive Systems Conference}}},
  author = {Lucero, Andr{\'e}s},
  year = {2012},
  month = jun,
  series = {{{DIS}} '12},
  pages = {438--447},
  publisher = {{Association for Computing Machinery}},
  address = {{New York, NY, USA}},
  doi = {10.1145/2317956.2318021},
  urldate = {2023-02-09},
  isbn = {978-1-4503-1210-3}
}

@inproceedings{palani2022don,
  title={” I don’t want to feel like I’m working in a 1960s factory”: The Practitioner Perspective on Creativity Support Tool Adoption},
  author={Palani, Srishti and Ledo, David and Fitzmaurice, George and Anderson, Fraser},
  booktitle={Proceedings of the 2022 CHI Conference on Human Factors in Computing Systems},
  pages={1--18},
  year={2022}
}

@book{greenbergSketchingUserExperiences2011,
  title = {Sketching User Experiences: {{The}} Workbook},
  shorttitle = {Sketching User Experiences},
  author = {Greenberg, Saul and Carpendale, Sheelagh and Marquardt, Nicolai and Buxton, Bill},
  year = {2011},
  publisher = {{Elsevier}}
}

@book{charmaz2006constructing,
  title={Constructing grounded theory: A practical guide through qualitative analysis},
  author={Charmaz, Kathy},
  year={2006},
  publisher={sage}
}

@inproceedings{zimmerman2007research,
  title={Research through design as a method for interaction design research in HCI},
  author={Zimmerman, John and Forlizzi, Jodi and Evenson, Shelley},
  booktitle={Proceedings of the SIGCHI conference on Human factors in computing systems},
  pages={493--502},
  year={2007}
}

@incollection{stappers2017research,
  title={Research through design},
  author={Stappers, Pieter Jan and Giaccardi, Elisa},
  booktitle={The encyclopedia of human-computer interaction},
  pages={1--94},
  year={2017},
  publisher={The Interaction Design Foundation}
}

\appendix
\clearpage
\section{Questionnaire}
\label{sec:questionnaire}

To understand how participants perceived the role and qualities of generative systems within their creative workflows, we asked them to rate a set of statements on a 5-point Likert scale (Strongly Disagree → Strongly Agree). Below we provide the full wording of each questionnaire item alongside the label used in the paper.

\begin{enumerate}
    \item \textbf{Ownership}: I feel ownership of what I create with \_\_\_\_.
    \item \textbf{Agency}: I feel control and agency using \_\_\_\_.
    \item \textbf{Early Stage Thinking}: \_\_\_\_ supports my early stage thinking.
    \item \textbf{Late Stage Thinking}: \_\_\_\_ supports my late stage thinking.
    \item \textbf{Inspiration}: I use \_\_\_\_ for inspiration.
    \item \textbf{Final Product}: Creations from \_\_\_\_ could show up in a final product.
    \item \textbf{Malleability}: \_\_\_\_ feels malleable.
    \item \textbf{Integration to Process}: \_\_\_\_ integrates smoothly into my creative process.
    \item \textbf{Integration to Tools}: \_\_\_\_ integrates smoothly with my other tools.
    \item \textbf{End-to-End Support}: \_\_\_\_ supports end-to-end workflows.
    \item \textbf{Reflects Intent}: I feel confident that \_\_\_\_ will reflect my design intent.
    \item \textbf{Enjoyment}: I enjoy using \_\_\_\_.
    \item \textbf{Fun}: \_\_\_\_ feels fun to use.
    \item \textbf{Professional-feeling}: \_\_\_\_ feels professional grade.
    \item \textbf{Ability to Fix Outputs}: When I get an undesired output with \_\_\_\_, I feel I can easily fix it to get what I want.
    \item \textbf{Expressiveness}: \_\_\_\_ feels expressive.
    \item \textbf{Seen as Tool}: I see \_\_\_\_ as tools.
    \item \textbf{Seen as Collaborator}: I see \_\_\_\_ as collaborators.
    \item \textbf{Empowerment}: \_\_\_\_ makes me feel empowered.
    \item \textbf{Creativity}: \_\_\_\_ makes me feel creative.
\end{enumerate}


\end{document}